\documentclass[aps,prd,twocolumn,groupedaddress]{revtex4}

\usepackage{graphicx}

\usepackage{dcolumn}

\usepackage{bm}

\renewcommand\({\left(}
\renewcommand\){\right)}
\renewcommand\[{\left[}
\renewcommand\]{\right]}

\newcommand{\ra}{\rightarrow}

\def\lsim{\raise 0.4ex\hbox{$<$}\kern -0.8em\lower 0.62
ex\hbox{$\sim$}}

\def\gsim{\raise 0.4ex\hbox{$>$}\kern -0.7em\lower 0.62
ex\hbox{$\sim$}}

\def\lbar{{\hbox{$\lambda$}\kern -0.7em\raise 0.6ex
\hbox{$-$}}}

\newcommand\eq[1]{eq.~(\ref{#1})}
\newcommand\eqs[2]{eqs.~(\ref{#1}) and (\ref{#2})}
\newcommand\Eq[1]{Equation~(\ref{#1})}

\newcommand\pa{\partial}
\newcommand\p{\partial}

\newcommand\ee{\end{equation}}
\newcommand\be{\begin{equation}}
\def\bea{\begin{array}}
\def\eea{\end{array}}\def\ea{\end{array}}
\newcommand\ees{\end{eqnarray}}
\newcommand\bees{\begin{eqnarray}}
\def\nn{\nonumber}

\def\D{\Delta}
\def\a{\alpha}
\def\b{\beta}

\def\s{\sigma}
\def\g{\gamma}

\def\d{\delta}

\def\eps{\epsilon}

\def\dslash{\hspace{-1mm}\not{\hbox{\kern-2pt $\partial$}}}
\def\Dslash{\not{\hbox{\kern-4pt $D$}}}
\def\pslash{\not{\hbox{\kern-2.1pt $p$}}}
\def\kslash{\not{\hbox{\kern-2.3pt $k$}}}
\def\qslash{\not{\hbox{\kern-2.3pt $q$}}}

\newcommand{\vac}{|0\rangle}
\newcommand{\cav}{\langle 0|}

\def\p1{{\bf p}_1}
\def\p2{{\bf p}_2}
\def\k1{{\bf k}_1}
\def\k2{{\bf k}_2}

\newcommand{\emn}{\eta_{\mu\nu}}

\newcommand{\gmn}{g_{\mu\nu}}

\newcommand{\gRS}{g^{\rho\sigma}}

\newcommand{\hij}{h_{ij}}

\newcommand{\pam}{\pa_{\mu}}

\newcommand{\pan}{\pa_{\nu}}

\newcommand{\Tmn}{T_{\mu\nu}}

\newcommand{\dddM}{\kern 0.2em \raise 1.9ex\hbox{$...$}\kern -1.0em \hbox{$M$}}
\newcommand{\dddQ}{\kern 0.2em \raise 1.9ex\hbox{$...$}\kern -1.0em \hbox{$Q$}}
\newcommand{\dddI}{\kern 0.2em \raise 1.9ex\hbox{$...$}\kern -1.0em\hbox{$I$}}
\newcommand{\dddJ}{\kern 0.2em \raise 1.9ex\hbox{$...$}\kern-1.0em
\hbox{$J$}}
\newcommand{\dddcalJ}{\kern 0.2em \raise 1.9ex\hbox{$...$}\kern-1.0em
\hbox{${\cal J}$}}

\newcommand{\dddO}{\kern 0.2em \raise 1.9ex\hbox{$...$}\kern -1.0em
\hbox{${\cal O}$}}
\def\dddz{\raise 1.5ex\hbox{$...$}\kern -0.8em \hbox{$z$}}
\def\dddd{\raise 1.8ex\hbox{$...$}\kern -0.8em \hbox{$d$}}
\def\dddbd{\raise 1.8ex\hbox{$...$}\kern -0.8em \hbox{${\bf d}$}}
\def\ddbd{\raise 1.8ex\hbox{$..$}\kern -0.8em \hbox{${\bf d}$}}
\def\dddx{\raise 1.6ex\hbox{$...$}\kern -0.8em \hbox{$x$}}

\newcommand{\Sch}{Schwarzschild }

\newcommand{\mpl}{M_{\rm Pl}}

\newcommand{\lc}{\Lambda_c}

\begin{document}

\title{Zero-point quantum fluctuations and dark energy}

\author{Michele Maggiore}

\affiliation{D\'epartement de Physique Th\'eorique, Universit\'e de 
	     Gen\`eve, CH-1211 Geneva, Switzerland}

\begin{abstract}
In  the  Hamiltonian formulation of General Relativity the energy associated to an asymptotically flat  space-time 
with metric $\gmn$ is related to the Hamiltonian $H_{\rm GR}$ by
$E=H_{\rm GR}[\gmn]-H_{\rm GR}[\emn]$, where the subtraction of the flat-space contribution is necessary to get rid of an otherwise divergent boundary term. This classic result indicates that 
the energy associated to flat space does not gravitate. 
We apply the same principle to study the effect of zero-point fluctuations of quantum fields in cosmology, proposing that their contribution to the cosmic expansion is obtained computing 
the vacuum energy of quantum fields in a FRW space-time with Hubble parameter $H(t)$ and subtracting from it the flat-space contribution.
Then the term proportional to $\lc^4$ (where $\lc$ is the UV cutoff) cancels
and  the remaining (bare) value of the vacuum energy density 
is proportional to $\Lambda_c^2 H^2(t)$. After renormalization, this produces
a renormalized  vacuum energy density $\sim M^2H^2(t)$, where $M$ is the scale where
quantum gravity sets is, so for $M$ of order of the Planck mass a vacuum energy density
of the order of the critical density can  be obtained without any fine tuning. The counterterms can be chosen so that
the renormalized energy density and pressure satisfy
$p=w\rho$, with $w$ a parameter that can be fixed by comparison to the observed value, so in particular one can chose $w=-1$.
An energy density evolving in time as $H^2(t)$  is however observationally
excluded as an explanation for the dominant dark energy component which is responsible for the observed acceleration of the universe. We rather propose that  zero-point vacuum fluctuations  provide a new subdominant  ``dark" contribution to the
cosmic expansion that, for a UV scale  $M$ slightly smaller than  the Planck mass, is consistent
with existing limits and potentially detectable. 

\end{abstract}





\maketitle

\section{Introduction} 

The cosmological constant problem is a basic issue of modern 
cosmology (see e.g. \cite {Weinberg:1988cp,Peebles:2002gy,Padmanabhan:2002ji,Copeland:2006wr}). 
One aspect of the problem is
to understand
what is the effect of the zero-point fluctuations of quantum fields on the
cosmic expansion. 
Zero-point quantum fluctuations seem to give
a contribution	to the vacuum energy density of order $\Lambda_c^4$, where
$\lc$ is the UV cutoff.  Even for a cutoff as low as 
$\lc\sim 1\, {\rm TeV}$, corresponding to scales where quantum field theory is
well tested, this exceeds by many orders of magnitude the value of the critical density
of  the universe, which
is of order $(10^{-3}\, {\rm eV})^4$.
This aspect of the problem has by now a long history, which even goes back to
works of Nernst and of Pauli in the 1920s~\cite{Peebles:2002gy}.

It is clear that this $\Lambda_c^4$ contribution should somehow be subtracted.
Renormalization in quantum field theory (QFT) offers a logically viable
possibility.
One should not forget, in fact, that this cutoff-dependent contribution, as any
similar quantity in QFT, is just a bare quantity rather than  a physical
observable. 
Working for instance with a cutoff $\lc$ in momentum space one finds that, in Minkowski spacetime, the bare, cut-off dependent, contribution to the vacuum energy density is
\be\label{bare}
\rho_{\rm bare}(\lc)=c\lc^4\, ,
\ee
with $c$ some constant that depends on the number and type of fields  of the theory.
The standard procedure in QFT is to add to it a cutoff-dependent
counterterm, chosen so that it cancels the divergent part and leaves us with
the desired finite part (for an elementary discussion of this point in the
cosmological constant case, 
see e.g. the textbook~\cite{Maggiore:2005qv}), i.e.
\be\label{rhocount}
\rho_{\rm count}(\lc)=-c\lc^4+\rho_{\rm vac}\, ,
\ee
where $\rho_{\rm vac}\sim (10^{-3}\rm eV)^4$ is independent of the cutoff and equal to the observed value of the vacuum energy density (assuming that vacuum energy is indeed responsible for the observed acceleration of the universe).
The physical, renormalized, vacuum energy density
\be\label{eq3v3b}
\rho_{\rm ren}\equiv \rho_{\rm bare}(\lc)+\rho_{\rm count}(\lc)
\ee
is therefore by definition  independent of the cutoff and equal to the observed value
$\rho_{\rm vac}$. In principle this procedure is not different from what is done when one renormalizes, say, the electron mass or the electron charge in QED, where again with suitable counterterms one removes the divergent, cut-off dependent, parts and fixes the finite parts so that they agree with the observed values (in this sense the statement, often heard, that QFT gives a
{\em wrong} prediction for the cosmological constant, is not correct. Strictly speaking QFT
makes no prediction for the cosmological constant, just as it does not
predict the electron mass nor the fine structure constant).

Still, for the vacuum energy this
procedure is not really satisfying. The source of uneasiness  is partly due to the fact that
the counterterm must be fine tuned to a huge precision in order to cancel the
$\Lambda_c^4$ term and leave us with a much smaller result. In fact, if in
\eq{bare} we take $\lc$ at least larger than a few TeV, where  quantum field theory has been successfully tested,  $\rho_{\rm bare}$
is at least of order $({\rm TeV})^4=(\rm 10^{12}\, eV)^4$, so 
$\rho_{\rm count}$ in \eq{eq3v3b} must be fine tuned so that it cancels something of order
$(\rm 10^{12}\, eV)^4$, leaving a result of order $(\rm 10^{-3}\, eV)^4$. This fine tuning
becomes even much worse if one dares to take $\lc$ of the order of the Planck mass $\mpl\sim 10^{19}$~GeV.
The fact that the counterterm must be tuned to such a precision creates a naturalness problem.
Here, however,  one might 
argue that neither the bare term  nor the counterterm  have any physical
meaning and only their sum is physical, so this fine tuning is  different
from an unplausible cancellation between observable quantities. The same kind
of cancellation appears, for instance, when one computes the Casimir effect.
The crucial point, however, is that this renormalization procedure leaves us with no
clue as to the physical value that emerges from this cancellation, so it gives
no explanation of why the energy density associated to the cosmological
constant appears to have just a value of the order of the critical density of
the universe at the present 
epoch.

The point of view that we develop in this paper is that, even if renormalization must be an ingredient for understanding the physical effects of vacuum fluctuations, it is not the end of the story.
The Casimir effect mentioned above gives indeed a first hint of what could be the missing ingredient for a correct treatment of vacuum energies in cosmology. In the Casimir effect the quantity that gives rise to observable effects,  which have indeed been detected experimentally, is the difference between the vacuum energy in a given geometry (e.g., for the electromagnetic field, between two parallel conducting plates) and the vacuum energy computed in  a reference geometry, which is just  flat space-time in an infinite volume. Both terms are separately divergent as $\lc^4$, but their difference is finite and observable. This  can suggest that, to obtain the physical effect of the vacuum energy density on the expansion of the universe, one should similarly compute the vacuum energy density in a FRW space-time and subtract from it the value of a reference space-time, which is naturally taken as Minkowski space.

A possible objection to this procedure could be that General Relativity requires that any form of energy should be a source for the gravitational field, which seems to imply that even the vacuum energy associated to flat space should contribute.
A more careful look at the formalism of GR shows however that the issue is not so simple and that, in a sense, the subtraction that we are advocating
is in fact already part of the standard tenets of classical GR.
To define carefully the energy associated to a field configuration in GR, it is convenient to 
use the Hamiltonian formulation, which goes back to the classic paper by 
Arnowitt,  Deser and Misner
\cite{ADM} (ADM). As we will review in more detail in Section~\ref{sect:ADM},
in order to define the Hamiltonian of GR one must at first work in a finite three-dimensional volume, and then the Hamiltonian takes the form
\be\label{HHH}
H_{\rm GR}=H_{\rm bulk}+H_{\rm boundary}\, ,
\ee
where $H_{\rm bulk}$ is given by an integral over the three-dimensional finite spatial volume at fixed time, and 
$H_{\rm boundary}$ by an integral over its (two-dimensional) boundary. At this point one would like to define the energy of a classical field configuration as the value of this Hamiltonian, evaluated on the classical solution, but one finds both a surprise and a difficulty. The ``surprise" (which actually is just a consequence of the invariance under diffeomorphisms) is that 
$H_{\rm bulk}$, evaluated on any classical solution of the equations of motion, vanishes, so the whole contribution comes from the boundary term. The difficulty is that the boundary term, evaluated on any asymptotically flat metric (including flat space-time) diverges when the boundary is taken to infinity. The solution proposed by ADM is to subtract from this boundary term, evaluated on an asymptotically flat space-time with metric $\gmn$,
the same term computed in flat space-time. The corresponding energy (or mass) is finite and is known as the ADM mass, and provides the standard definition of mass in GR.
For instance, when applied to the \Sch space-time, the ADM mass computed in this way turns out to be equal to the mass  $M$ that appears in the \Sch metric.

The ADM prescription can be summarized by saying that, in GR, the energy $E$ 
associated to an asymptotically flat space-time with metric
$\gmn$ can be obtained from the Hamiltonian $H_{\rm GR}$  by
\be\label{EHH}
E=H_{\rm GR}[\gmn]-H_{\rm GR}[\emn]\, ,
\ee
where $\emn$ is the flat metric. Even if the context in which this formula is valid, namely asymptotically flat space-times, is  different from the cosmological context in which we are interested here, still \eq{EHH} suffices to make the point that the intuitive idea that GR requires that any form of energy acts as a source for the gravitational field is not really correct. \Eq{EHH} tells us that the energy associated to Minkowski space does not gravitate.

The idea of this paper is to generalize \eq{EHH} to the case of zero-point quantum fluctuation in curved space, proposing that the effect of zero-point fluctuations  on the cosmic expansion should be obtained by computing 
the vacuum energy of quantum fields in a FRW space-time with Hubble parameter $H(t)$ and subtracting from it the flat-space contribution. Computed in this way, the physical effect of zero-point fluctuations on the cosmic expansion can be seen as a sort of ``cosmological Casimir effect": while in the standard Casimir effect one computes the vacuum energy in a given geometry (say, for the electromagnetic field, between two infinite parallel conducting planes) and subtracts from it the value computed in  a reference geometry (flat space-time in  an infinite volume), here we compute the vacuum energies of the various fields in a given curved space-time, e.g. in FRW 
space-time, and we subtract from it the value computed in a reference space-time, i.e. Minkowski.

A possible objection to the idea of applying \eq{EHH} to vacuum fluctuations could be that one might think that, after all, the structure of UV divergences is determined by local properties of the theory, while whether a space-time is Minkowski is a global question. However
one should keep in mind that, even if $\Tmn(x)$ 
is a local quantity,
the vacuum expectation value $\cav\Tmn\vac$ in a curved background  involves the global aspects of the space-time in which it is computed.
This is due to the fact  that 
$\cav\Tmn\vac$ requires a definition of the vacuum state $\vac$. 
The vacuum is  defined from the condition $a_{\bf k}\vac=0$, where 
the annihilation operators $a_{\bf k}$ are defined with respect to a set of mode functions $\phi_{\bf k}(t)$. The mode functions, in turn, are obtained solving a wave equation over the whole space-time, and therefore are sensitive to global properties of the space-time itself. In particular,
 in a FRW background the mode functions   depend on the scale factor so their time derivatives (which enter in the computation of $\cav\Tmn\vac$) depend on the Hubble parameter $H(t)$. As we will review below,  this results in the fact that  the quadratic divergence in $\cav\Tmn\vac$ depends on the expansion rate $H(t)$ of the FRW background.

The paper is organized as follows. 
In Section~\ref{sect:ADM} we recall 
how \eq{EHH} is derived in the ADM formalism.
The reader familiar with the subject, or not interested in the derivation, might simply wish to move directly to Section~\ref{sect:appl}, where we apply this classical subtraction, together with standard renormalization theory, to zero-point vacuum fluctuations. Some cosmological consequences of our proposal are discussed in Section~\ref{sect:impl}, while Section~\ref{sect:Concl} contains our conclusions.

\section{Subtractions in classical GR: the ADM mass}\label{sect:ADM}

In this section we briefly discuss how the Arnowitt-Deser-Misner
(ADM) mass  is defined in GR
\cite{ADM} (we follow the very clear discussion of the textbook~\cite{pois04}). We
begin by recalling that, in a finite four-dimensional volume 
${\cal V}$ with boundary $\pa {\cal V}$, the gravitational action is (setting
$c=1$ and using the signature 
$\emn=(-,+,+,+)$)
\be\label{3SEbound}
S_{\rm grav}=\frac{1}{16\pi G}\, \int_{\cal V} d^4x\, \sqrt{-g}\, R
+\frac{1}{8\pi G}\, \int_{\pa\cal V} d^3y\,  \sqrt{|h|}\eps K\, ,
\ee
where $\gmn$ is the four-dimensional metric, $\hij$ is the metric induced on
the boundary  
$\pa {\cal V}$, $h=\det \hij$,
$K$ is the trace of the extrinsic curvature of the boundary,  $y^i$ are the
coordinates of the boundary, 
and $\eps=+1$ on the regions of the boundary 
where ${\pa\cal V}$ is time-like and $\eps=-1$ where ${\pa\cal V}$ is
space-like.
The first term is the usual Einstein-Hilbert action, while the second is a
boundary term which is necessary to obtain a well-defined variational
principle.

To pass to the Hamiltonian formalism one performs the $3+1$ decomposition of
the metric,
\be\label{3metricADM}
ds^2=-\a^2 dt^2 +h_{ij} (dx^i+\b^idt)(dx^j+\b^jdt)\, ,
\ee
where $\a$ and $\b^i$ are the lapse function and shift vector, respectively,
and $\hij$ is the induced 
metric on the 3-dimensional spatial hypersurfaces. One defines as usual the
conjugate momentum
$\pi^{ij}={\pa {\cal L}}/{\pa\dot{h}_{ij}}$,  where ${\cal L}$ is the ``volume
part'' of the Lagrangian density,
and the Hamiltonian  $H$ is then the volume integral of  the Hamiltonian density 
${\cal H}=\pi^{ij}\dot{h}_{ij}-{\cal L}$.  The
explicit computation 
(see e.g. Section 4.2.6 of  \cite{pois04}) gives
\bees\label{3calHADM} 
(16\pi G)\,  H&=&\int_{\Sigma_t} d^3x\sqrt{h}
\( -\a C_0-2\b_iC^i \)\\
&&-2\int_{S_t} d^2\theta\sqrt{\s}
\[\a k-\beta_i r_j(K^{ij}-K h^{ij})\]\, ,\nn
\ees
where $\Sigma_t$ denotes the three-dimensional spatial hypersurfaces at fixed
$t$ and $S_t$ (with coordinates $\theta^i$) is the intersection of $\Sigma_t$
with $\pa{\cal V}$. In an asymptotically flat space-time $S_t$ is just a
2-sphere at large radius $r=R$ and fixed $t$;
$C_0$ and $C^i$  depend only on $\hij$ and on its derivatives, but not on $\a$ and $\b_i$, while
$k$ in \eq{3calHADM} is the trace of the extrinsic
curvature of $S_t$, and $\s$ is the determinant of the two-dimensional induced
metric.

One would like to define the energy of a classical field configuration, i.e. of
a solution of the equations of motion, as the value of this Hamiltonian on the
solution.
Performing the variation with respect to $\a$ and $\b_i$ one obtains the
constraint equations $C_0=0$ and $C^i=0$. Therefore, on a classical solution,
the volume term in \eq{3calHADM} vanishes, and only the boundary term
contributes. The ADM mass is defined by setting $\a=1$ and $\b_i=0$
after performing the variation (corresponding to the fact that energy is associated to asymptotic  time translations; setting $\a=0$ and $\b^i=1$ one rather obtains the ADM momentum $P^i_{\rm ADM}$),
so only the term proportional to $k$  contributes to the mass.
However, even for Minkowski space, this
boundary term diverges. In fact, for an asymptotically flat space-time we can
take
${\pa\cal V}$ to be a three-dimensional cylinder made of the two
three-dimensional 
time-like hypersurfaces $\{t=t_1, r\leq R\}$ 
and $\{t=t_2, r\leq R\}$
(the ``faces" of the three-dimensional cylinder) and of the space-like
hypersurface 
$\{r=R, t_1\leq t\leq t_2\}$. Let us denote by $K_0$ the extrinsic curvature of
$\pa{\cal V}$ computed with a flat Minkowski metric. 
The faces at $t=t_1$ and $t_2$ have $K_0=0$; however, on the surface
$\{r=R, t_1\leq t\leq t_2\}$ the extrinsic 
curvature is $K_0=2/R$, while 
$|h|^{1/2}=R^2\sin^2\theta$ and $\eps=+1$, so the boundary term in
\eq{3metricADM} is~\cite{pois04}
\be
\int_{\pa\cal V} d^3y\,  \sqrt{|h|}\eps K_0=8\pi (t_2-t_1) R\, ,
\ee
and diverges both when	$R\ra\infty$ and when $(t_2-t_1)\ra\infty$. As a
consequence, also the boundary term proportional to $k$ in \eq{3calHADM}
diverges, already for the Minkowski metric, and then of course also for generic
asymptotically flat space-times.
The ADM prescription is then to replace $K$ in \eq{3SEbound} by $(K-K_0)$, i.e.
to subtract from the trace of the extrinsic curvature computed with the desired
metric, the value computed in flat space. Correspondingly, the trace $k$ of the
two-dimensional extrinsic curvature in \eq{3calHADM} becomes $(k-k_0)$, and the
ADM mass of an asymptotically flat space-time is defined as~\cite{ADM}
\be
M_{\rm ADM}=-\frac{1}{8\pi G}\lim_{S_t\ra\infty}\int_{S_t} d^2\theta\sqrt{\s}
(k-k_0)\, .
\ee
If for instance one applies this definition to the \Sch space-time, one finds
that $M_{\rm ADM}$ is equal to the mass $M$ which appears in the \Sch metric.

What we learn from this is that, in classical GR, the mass or the energy that
acts as the source of curvature of space-time, such as for instance the mass
$M$ that enters in the \Sch metric, can be obtained from a Hamiltonian
treatment only after subtracting a flat-space contribution that need not be
zero, and is in fact  even divergent, but still does not act as a source of
curvature.

\section{Application to zero-point energies}\label{sect:appl}

It is natural to apply the same principle to zero-point quantum fluctuations.
In particular, in a cosmological setting, we propose that the zero-point energy
density and pressure that contribute to the cosmological expansion are obtained
by computing
the energy density and pressure due to 
zero-point quantum fluctuations
in a FRW metric   with Hubble parameter $H(t)=\dot{a}/a$,
and subtracting from it the flat-space contribution.

We consider first the contribution of a real massless scalar field.
In a FRW background  the mode expansion of the field is
\be\label{phi}
\phi(x)=
\int\frac{d^3k}{(2\pi)^3 \sqrt{2k}}\, 
\[
a_{\bf k}\phi_{k}(t) e^{i{\bf k\cdot x}}+
a_{\bf k}^{\dagger}\phi^*_{k}(t) e^{-i{\bf k\cdot x}}\]\, ,
\ee
where ${\bf k}$ is the comoving momentum, and $\phi_k(t)$ satisfies the
massless Klein-Gordon equation in a FRW background, 
\be\label{eqforphi}
\phi_k''+2\frac{a'}{a}\phi_k'+k^2\phi_k=0\, ,
\ee
where the prime is the derivative with respect to conformal time $\eta$.
Writing $\phi_k=\psi_k/a$ this equation is reduced to the  form
\be\label{eqpsik}
\psi_k''+\(k^2-\frac{a''}{a}\)\psi_k=0\, .
\ee
In a De~Sitter background we have $a(\eta)=-1/(H\eta )$ so $a''/a=2/\eta^2$,
while in a matter-dominated (MD) epoch
$a\sim\eta^2$ and therefore again
$a''/a=2/\eta^2$. Thus, in both cases the positive frequency solution of
\eq{eqpsik} is 
\be
\psi_k(\eta)=\(1-\frac{i}{k\eta}\) e^{-ik\eta}\,  ,
\ee
and therefore $\phi_k(\eta)$ is given by
\be\label{phiketa}
\phi_k(\eta)=\frac{1}{a(\eta)}\(1-\frac{i}{k\eta}\) e^{-ik\eta}\, ,
\ee
with $a(\eta)$ the scale factor of De~Sitter or MD epoch, respectively.
In contrast, during a radiation-dominated (RD) epoch
$a\sim\eta$, so
$a''/a=0$. Then $\psi_k(\eta)=e^{-ik\eta}$ and
\be\label{modesRD}
\phi_k(\eta)=\frac{1}{a(\eta)}e^{-ik\eta}\,  .
\ee
Using the
energy--momentum tensor of a minimally coupled massless scalar field, 
\be\label{Tmunu}
T_{\mu\nu}=\pam\phi \pan\phi-\frac{1}{2}\gmn
\gRS\pa_{\rho}\phi\pa_{\sigma}\phi\, ,
\ee
and setting $\gmn=(-1,a^2\delta_{ij})$, a simple computation \cite{Parker:1974qw,Fulling:1974zr}, reviewed in
Appendix~\ref{app:A},  shows that
the off-diagonal elements of $\cav\Tmn\vac$ vanish, while
the  vacuum energy density and pressure are given by
\bees
\rho &=&\frac{1}{2}\int\frac{d^3k}{(2\pi)^3 2k}\, 
\( |\dot{\phi}_k|^2+\frac{k^2}{a^2}|\phi_k|^2\)\, ,\label{rhophikd3k}\\
p&=&\frac{1}{2}\int\frac{d^3k}{(2\pi)^3 2k}\, 
\( |\dot{\phi}_k|^2-\frac{k^2}{3a^2}|\phi_k|^2\)\, ,\label{pphikd3k}
\ees
where the dot denotes  the derivative with respect to cosmic time $t$.  As they stand, these expressions must still be regularized, and we regularize them by putting a cutoff in momentum space. Recall that  the comoving momentum
$k$ is just a  label  of the Fourier mode under consideration, while the
physical momentum of the mode
is given by  $k/a$. We expect that quantum gravity enters the game when the
{\em physical} momenta exceed the Planck scale, and we therefore put a 
time-independent cutoff $\lc$ over physical, rather than comoving, momenta. In terms of comoving momentum $k$ this means $k< a(t)\lc$. 
Since the modes $\phi_k$ depend only on 
$k=|{\bf k}|$, the angular integrals are trivially performed, and we finally get
\bees
\rho &=&\frac{1}{8\pi^2}\int_0^{a\lc} dk\, k
\( |\dot{\phi}_k|^2+\frac{k^2}{a^2}|\phi_k|^2\)\, ,\label{rhophik}\\
p&=&\frac{1}{8\pi^2}\int_0^{a\lc} dk\, k
\( |\dot{\phi}_k|^2-\frac{k^2}{3a^2}|\phi_k|^2\)\, .\label{pphik}
\ees

\subsection{Vacuum fluctuations in De~Sitter space}

\subsubsection{Vacuum energy density}

We compute first these expressions in De~Sitter space. 
We therefore plug \eq{phiketa}, with  $a(\eta)=-1/(H\eta)$,  into
\eqs{rhophik}{pphik}. For the energy density the result is
\bees
\rho(\lc)&=&\frac{1}{4\pi^2}\int_0^{a\lc}dk\,  k
\(\frac{k^2}{a^4}+\frac{H^2}{2a^2}\)\, \nn\\
&=&\frac{\lc^4}{16\pi^2}+\frac{H^2\lc^2}{16\pi^2}\, .\label{rhovac1}
\ees
The first term is the well known flat-space result, proportional to the fourth
power of the cutoff, while the term quadratic in $\lc$ is the correction due  
to the expansion of the universe. The appearance of a term $\sim H^2\lc^2$,  
in a theory with two 
scales, the UV cutoff $\lc$ and the Hubble scale $H$,
can also be understood using rather general 
arguments~\cite{Padmanabhan:2004qc}. 
 
According to our proposal we now subtract the flat-space contribution, which is
simply the term
$\sim \lc^4$ in \eq{rhovac1}, and we find that in De~Sitter space the
zero-point quantum fluctuations of a real
scalar field have a (bare) energy density
\be\label{rhoZ}
\rho_{\rm bare} (\lc)=\frac{H^2\lc^2}{16\pi^2}\, .
\ee
The subscript ``bare" stresses  that this is still  a bare, cut-off dependent quantity. We have eliminated the quartic divergence thanks to the ``classical'' prescription (\ref{EHH}), but the result is still a bare energy density, diverging quadratically with the cutoff. In this sense, the situation is different from the usual Casimir effect, where the subtraction of the flat-space contribution suffices to make the result finite. However, we can now  use standard renormalization theory, so the renormalized energy density is obtained by adding a counterterm whose divergent part is chosen so to cancel this quadratic divergence, and whose finite part $\rho_{\rm vac}$  is in principle fixed by the observation, so
\be\label{rhoZcount}
\rho_{\rm count} (\lc)=-\frac{H^2\lc^2}{16\pi^2}+\rho_{\rm vac}\, .
\ee
As usual in the renormalization procedure, the finite part $\rho_{\rm vac}$ cannot be predicted. It must be fixed to the observed value. What we have gained, with respect to \eq{rhocount}, is that now we have a different understanding of what is a ``natural" value of this finite part. 
If quantum gravity sets in at a mass scale $M$ (so that $M$ could be typically given by the Planck mass $\mpl$, or by the string mass), a natural, non fine-tuned value of 
$\rho_{\rm vac}$ is given by
\be\label{rhonat}
\rho_{\rm vac}= {\rm const.}\times \s\frac{H^2M^2}{16\pi^2}\, ,
\ee
where  ``${\rm const.}$" is a constant ${\cal O}(1)$, which cannot be fixed by naturalness arguments only, and must be determined by comparison with the experiment.\footnote{More precisely, renormalization  trades the cutoff $\lc$ for the subtraction point $\mu$. While the latter is in principle arbitrary, a clever choice of $\mu$ will minimize the effect of radiative corrections. For instance, if one renormalizes the electroweak theory in the MS scheme, one naturally takes $\mu=m_W$. Using a much lower  subtraction point, say $\mu=m_e$, is in principle legitimate, but all radiative corrections would then become large, and the whole perturbative approach could be spoiled. In this sense, $\mu=m_W$ is a natural subtraction point for the electroweak theory, and we similarly expect that, in a theory involving quantum gravity, the natural subtraction point will be given by the Planck or string mass. (I thank a referee for this comment).}

Observe that  naturalness arguments cannot fix the sign of the finite part either, and we have added a factor
$\s=\pm 1$ to take this fact explicitly into account. There is in fact no a priori reason why the renormalized vacuum energy should necessarily be positive. For instance, the vacuum energies obtained for fields in a finite volume from the Casimir effect can be either positive or negative, depending e.g. on the type of field and on  the geometry considered.

\subsubsection{Equation of state and general covariance}

Repeating the same computation for the pressure we get
\bees
p(\lc)&=&\frac{1}{4\pi^2}\int_0^{a\lc}dk\,  k
\(\frac{k^2}{3a^4}-\frac{H^2}{6a^2}\)\nn\\
&=& \frac{\lc^4}{48\pi^2}-\frac{H^2\lc^2}{48\pi^2}
\, ,\label{pvac1}
\ees
The term $\sim\lc^4$ in \eq{pvac1} was already
computed in 
\cite{Akhmedov:2002ts} and is just the result
in Minkowski space, that we subtract, so we end up with
\be
p_{\rm bare}(\lc)=-\frac{H^2\lc^2}{48\pi^2}\, .
\ee
Observe that $p_{\rm bare}(\lc)=-(1/3)\rho_{\rm bare}(\lc)$. However, it would be incorrect to conclude that vacuum fluctuations in De~Sitter space satisfy the equation of state  $p=w\rho$ with $w=-1/3$. The point is that this relation only holds for the  bare quantities, and not necessarily for the renormalized ones. The physical, renormalized, pressure  is obtained by adding a counterterm
\be\label{pZcount}
p_{\rm count} (\lc)=+\frac{H^2\lc^2}{48\pi^2}+p_{\rm vac}\, .
\ee
Observe that regularizing the theory with a cutoff over spatial momenta, as we have done, breaks explicitly Lorentz invariance in Minkowski space,  since the notion of maximum value of spatial momenta is not invariant under boosts. In a generic FRW background, of course, Lorentz transformations are not a symmetry of the theory, since the metric depends explicitly on time, and the guiding principle is rather general covariance, which again is broken by a cutoff over spatial momenta.
However, even if the regularization breaks general covariance,  in De~Sitter space
a generally covariant result can still be obtained in the end for the physical, renormalized, vacuum expectation value of the energy-momentum tensor, just by choosing
the finite parts of the counterterms  such that $p_{\rm vac}=-\rho_{\rm vac}$. Then the vacuum expectation value of the renormalized energy-momentum tensor 
$T^{\mu}_{\nu}={\rm diag}(-\rho,p,p,p)$ becomes 
\be
\cav T^{\mu}_{\nu}\vac ={\rm const.} \times  \s\frac{H^2M^2}{16\pi^2} (-\d^{\mu}_{\nu})\, ,
\ee
or, lowering the upper index with the  metric $\gmn$,
\be\label{TmngmnDS}
\cav \Tmn\vac ={\rm const.}\times   \s\frac{H^2M^2}{16\pi^2} (-\gmn)\, .
\ee
Since in De~Sitter space $H$ is constant we see that, with the choice
$p_{\rm vac}=-\rho_{\rm vac}$, 
 $\cav \Tmn\vac$ is just given by a numerical constant times $\gmn$, and it is therefore covariantly conserved. In the language of the effective action for gravity, which is
obtained by treating the metric $\gmn$ as a classical background and integrating over the matter degrees of freedom
(see e.g. refs.~\cite{Barvinsky:1985an,Buchbinder:1992rb,Shapiro:2008sf}), the  vacuum expectation value of the energy-momentum tensor  is given by the functional derivative
$(2/\sqrt{-g})\d/\d\gmn$ of the effective action. Then a contribution such as that given in 
\eq {TmngmnDS} can be obtained by taking the functional derivative of the volume term in the effective action
(see ref.~\cite{Pelinson:2010yr} for a discussion of the equation of state that can be obtained from the various contributions to the effective action).

The choice $p_{\rm vac}=-\rho_{\rm vac}$ will therefore be assumed in the following, for De~Sitter space-time. Observe that a 
covariant result for the renormalized value of $\cav\Tmn\vac$  is obtained with a counterterm  that is not covariant, i.e. is not proportional to $\gmn$, 
since $p_{\rm count} (\lc)\neq -\rho_{\rm count} (\lc)$,
but again this is a consequence of the fact that our regularization is not covariant. 

\subsubsection{Contribution of higher-spin fields and massive particles}

A similar conclusion 
about the ``natural" value of the energy density of vacuum fluctuations  holds if we add the
contribution of fields with
different spin.
In fact, massless fermions and gauge bosons have a conformally  invariant action and then, since
the FRW metric is conformally equivalent to flat Minkoswki space, their vacuum
energy in a FRW space-time is the same as in Minkowski space.  Therefore, with our subtraction (\ref{EHH}),
they do not contribute to $\rho_{\rm vac}$ and $p_{\rm vac}$.
This is completely equivalent to the well known fact that vacuum
fluctuations of massless fermions and gauge bosons are not amplified during
inflation.

The contribution of massive fermions to vacuum fluctuations in De~Sitter space, as well as the generalization of
\eq{rhovac1} to massive bosons, can be readily computed.
For massive bosons, \eq{rhovac1} becomes \cite{Fulling:1974zr,Bilic:2010xd}
\be\label{rhovac1massive}
\rho_B(\lc)=\frac{1}{4\pi^2}\int_0^{a\lc}dk\,  k
\[\frac{k\omega_{k}}{a^3}+\frac{H^2 k}{2a^3\omega_k}\(1+\frac{m^2}{\omega_k^2}\)
\]\, ,
\ee
where we have neglected terms of order $H^4$, and terms convergent in the UV; $\omega_k$ is defined as
\be
\omega_k=\sqrt{m^2+(k/a)^2}\, ,
\ee
so in the massless limit $\omega_k\ra k/a$ and we recover \eq{rhovac1} \footnote{Observe that in 
ref.~\cite{Bilic:2010xd} the result is written directly for the {\em complex} scalar belonging to a  chiral superfield, hence the bosonic vacuum energy in eq.~(32) of ref.~\cite{Bilic:2010xd} is twice as large as that in  our \eq{rhovac1}, which holds for a {\em real} scalar field.}.
In this massive case our prescription amounts to subtracting the flat-space contribution given by the term $k\omega_k/a^3$ in brackets. The remaining, quadratically divergent term, leads again to 
\eq{rhonat}, times a correction  $[1+{\cal O}(m^2/M^2)]$, and to a logarithmically divergent term,  which after renormalization produces a contribution to $\rho_{\rm vac}$  proportional  
to $H^2m^2\log(M/m)$, and therefore subleading for $m\ll M$.

The contribution of a massive Majorana spinor field to the vacuum energy density in De~Sitter 
space is instead~\cite{Bilic:2010xd}
\be\label{rhovac1massiveferm}
\rho_F(\lc)=\frac{1}{2\pi^2}\int_0^{a\lc}dk\,  k
\[-\frac{k\omega_{k}}{a^3}+\frac{m^2H^2 k}{8a^3\omega^3_k}
\]\, .
\ee
The first term in brackets gives the usual negative contribution to vacuum energy in flat space due to fermions. Observe that in a supersymmetric model, where to each Majorana spinor is associated a complex scalar field, and therefore two real scalar fields, the contribution $(-k\omega_k/a^3)$ in
\eq{rhovac1massiveferm} cancels exactly the bosonic contribution  $(+k\omega_k/a^3)$
in \eq{rhovac1massive}, giving the usual cancellation of vacuum energy for a supersymmetric theory in Minkowski space. In a realistic model with  supersymmetry broken at a scale $\Lambda_{\rm susy}$, this cancellation is however only partial and leaves the usual result of order $\Lambda_{\rm susy}^4$. In our approach, instead, the quartic divergence are eliminated exactly by the  prescription of subtracting the flat-space contribution.

The term that remains in \eq{rhovac1massiveferm}, after subtraction of the flat-space contribution, gives a divergent contribution equal to $(m^2H^2/16\pi^2)\log(\lc/m)$
(consistently with the fact that, for $m=0$, it must vanish because of conformal invariance).
After renormalization, this gives a contribution to the vacuum energy which is of order
$(m^2H^2/16\pi^2)\log(M/m)$, and which therefore 
for $m\ll M$, is  negligible with respect to the bosonic contribution (\ref{rhonat}). It is also interesting to observe that, even in a theory with exact supersymmetry, the cancellation between fermionic and bosonic divergences only takes place at the level of the quartic divergence. There is no contribution proportional to $\lc^2$ from the fermionic sector, and therefore  the whole contribution
proportional to $H^2\lc^2$ comes from the bosonic sector. This means that 
\eq{rhonat} holds even in a theory with exact or broken supersymmetry.

In contrast, gravitons  give a contribution to the bare energy density proportional to $\lc^2H^2$. 
In fact, in a FRW space-time each of
the two helicity modes $h_{\a,k}$, with $\a=\{+,\times\}$ satisfies separately
the same wave equation as \eq{eqforphi} with $\phi_k$ replaced by $h_{\a,k}$,
\be
h''_{\a,k}+2\frac{a'}{a}h'_{\a,k}+k^2h_{\a,k}=0\, .
\ee
So each of
the two helicity modes gives the same contribution to $\rho_{\rm vac}$ and $p_{\rm vac}$ as a
minimally coupled massless scalar field. Therefore,  in a theory with $n_s$ minimally coupled elementary scalar fields plus the two degrees of freedom for the graviton, the natural value for the energy density, \eq{rhonat},  is of order
$(n_s+2)H^2M^2/(16\pi^2)$. In particular,  even in a theory with no fundamental scalar field, a contribution of order $H^2M^2/(8\pi^2)$ comes anyhow from gravitons.

\subsection{Vacuum fluctuations during RD and MD}\label{sect:RDMD}

It is straightforward to repeat the same calculation for a radiation-dominated 
(RD) and for  a matter-dominated (MD) era. 
For the RD epoch we use the modes (\ref{modesRD}).  Then, recalling that $dt=ad\eta$,
\be
\dot{\phi}_k=\frac{1}{a}\phi'_k=-\frac{1}{a^2}\[ik+\frac{a'}{a}\] e^{-ik\eta}\, ,
\ee
where as usual the dot denotes the derivative with respect to cosmic time $t$ and the prime the derivative with respect to $\eta$. Using $a'/a=\dot{a}=aH(t)$, we get
\bees
\rho(\lc)&=&\frac{1}{4\pi^2}\int_0^{a\lc}dk\,  k
\(\frac{k^2}{a^4}+\frac{H^2(t)}{2a^2}\)\, \nn\\
&=&\frac{\lc^4}{16\pi^2}+\frac{H^2(t)\lc^2}{16\pi^2}\, ,\label{rhovac1RD}
\ees
so the energy density  turns out to be identical to
\eq{rhovac1}, except that now $H$ is replaced by $H(t)$.
For the pressure we get
\be
p(\lc)=\frac{\lc^4}{48\pi^2}+\frac{H^2(t)\lc^2}{16\pi^2}\, .\label{pvac1RD}
\ee
Thus, once removed the Minkowski term, we remain with 
\be
\rho_{\rm bare}(\lc)=\frac{H^2(t)\lc^2}{16\pi^2}\, ,
\ee
while for the pressure we get
$p_{\rm bare}(\lc)=\rho_{\rm bare}(\lc)$. Similarly to what we have discussed in the de~Sitter case, this means that the natural value of the renormalized energy density is 
\be\label{rhoH2M2}
\rho_{\rm vac}(t)={\rm const.}\times\s\frac{H^2(t)M^2}{16\pi^2}\, ,
\ee
where $M$ is  the scale where quantum gravity sets in, ``${\rm const.}$" is a numerical 
constant ${\cal O}(1)$,  and $\s=\pm 1$. On the other hand, the relation $p_{\rm bare}(\lc)=\rho_{\rm bare}(\lc)$ does not imply that the renormalized energy density and the renormalized pressure satisfy an equation of state with $w=+1$. As in the de~Sitter case, the finite part in the counterterm for the pressure can be chosen so to reproduce any observed value of $w$, in particular the  value $w=-1$. 
The issue of general covariance is however now more subtle, since even the choice $w=-1$ now leads to
\be\label{TmngmnRD}
\cav \Tmn\vac ={\rm const.}\times   \s\frac{H^2(t)M^2}{16\pi^2} (-\gmn)\, 
\ee
which, because of the time dependence of $H(t)$, is no longer covariantly conserved. However, 
general covariance actually only requires that the {\em total} energy momentum tensor, including that of matter, radiation, etc., be covariantly conserved. The fact that energy-momentum tensor associated to vacuum fluctuations is not separately conserved means that there must be an energy exchange between vacuum fluctuations and other energy sources such as radiation or matter, so that we are actually dealing with an interacting dark energy model. We will come back to this issue in 
Section~\ref{sect:impl}.

For MD we use the modes given in \eq{phiketa}, with $a(\eta)\sim\eta^2$.
For the energy density we find
\be\label{rhoclcMD}
\rho(\lc)=\frac{1}{4\pi^2}\int_0^{a\lc}dk\,	k
\(\frac{k^2}{a^4}+\frac{H^2(t)}{2a^2}+\frac{9H^4(t)}{32 k^2}\)\, .
\ee
So, both for MD and for RD, the first two terms are the same as in 
De~Sitter, \eq{rhovac1}, except that $H$ becomes $H(t)$. Again, the first term
is the Minkowski contribution, that we subtract. The term $\sim H^4$ diverges
only logarithmically with the UV cutoff (and also require an IR cutoff, which for a light scalar field is provided by its mass, while for a strictly massless field
could be taken equal to $H$). Thus, for the physical, renormalized, energy density  we find again that the natural value is given by \eq{rhoH2M2},
since the term $\sim  H^4$ in \eq{rhoclcMD} produces a subleading term of order
$H^4(t)\log (M/H(t))$.

Recalling that (in units $\hbar=c=1$) Newton's constant is $G=1/\mpl^2$,
and that the critical density at time $t$ is 
\be
\rho_c(t)=\frac{3H^2(t)}{8\pi G}=\frac{3}{8\pi} H^2(t)\mpl^2\, ,
\ee
we
can express the above result by saying that the ``natural'' value suggested by QFT is
\be\label{rhoZ2}
\rho_ {\rm vac}(t)={\rm const.}\times \sigma \frac{1}{6\pi}\, \(\frac{M}{\mpl}\)^2\, \rho_c(t)\, ,
\ee
with ${\rm const.}={\cal O}(1)$ and $\s=\pm 1$.
We therefore find that, both during RD and MD, the renormalized energy density due to  zero-point quantum
fluctuations is not a constant, and therefore
does not contribute to the cosmological constant. Rather, it is a fixed
fraction of the critical density $\rho_c(t)$. 

As before, each helicity mode of a graviton contributes as a minimally coupled scalar field,
while gauge bosons do not contribute and the contribution of light fermions is suppressed by a factor $(m/M)^2\log(M/m)$. 
Thus, in a theory
with $n_s$ fundamental minimally coupled scalar fields plus the two helicity modes of
the graviton we have, for either De~Sitter, RD or MD,
\be\label{defOmegaZ}
\rho_Z(t)=\Omega_Z\rho_c(t)\, ,
\ee
where the subscript $Z$ stands for ``zero-point quantum fluctuations", and the natural value of $\Omega_Z$ is
\be\label{OmZ}
\Omega_Z\simeq\s
\frac{(n_s+2)}{6\pi}\, \(\frac{M}{\mpl}\)^2\,,
\ee
where $M$ is the scale where quantum gravity sets in, e.g.  $\mpl$ itself or the string scale.
As we have seen in \eqs{rhovac1massive}{rhovac1massiveferm}, bosons or fermions with mass $m$ not far from the quantum gravity scale $M$ would give corrections of order $[1+{\cal O}(m^2/M^2)]$ to this estimate (which could become numerically important if there were, e.g. many fermions at a scale, such as the GUT scale, not far from the Planck mass).
The computation of the pressure during MD  gives
\be
p(\lc)=\frac{1}{4\pi^2}\int_0^{a\lc}dk\,  k
\(\frac{k^2}{3a^4}+\frac{H^2(t)}{3a^2}+\frac{9H^4(t)}{32 k^2}\)\, .
\ee
Again, the first term is the Minkowski contribution, that we subtract, and we
neglect
the term $\sim H^4$. Thus, we finally find
\be
p_{\rm bare}(\lc)=\frac{H^2(t)\lc^2}{24\pi^2}\, ,
\ee
and therefore, during MD, $p_{\rm bare}=+(2/3)\rho_{\rm bare}$, but again the renormalized 
energy and pressure satisfy $p=w\rho$ with $w$ determined by the 
observation.\footnote{It is curious to observe that in all three case (De~Sitter, RD and MD) the 
bare quantities $p_{\rm bare}$ and $\rho_{\rm bare}$ satisfy
$p_{\rm bare}=w_{\rm bare}\rho_{\rm bare}$
with $w_{\rm bare}=(2/3)+w_{\rm dom}$ where $w_{\rm dom}$ is the $w$-parameter of the component that dominates the evolution during the corresponding phase, i.e.
$w_{\rm dom}=-1,+1/3,0$ during  De~Sitter, RD and MD,
respectively.}

A technical point that
deserves some comment is the choice of the modes given in
\eqs{phiketa}{modesRD}. These modes are particularly natural
since in the UV limit they reduce to positive-frequency plane
waves in flat space. However, the choice of the modes is equivalent to the choice
of a particular vacuum state, and the most general possibility is
a superposition of positive- and negative-frequency modes with
Bogoliubov coefficients $\a_{k}$ and $\b_{k}$.
As we show in appendix
\ref{app:B}, for a generic choice of vacuum the dependence of the natural value of $\rho_{\rm vac}$ on $H^2(t)M^2$ is not altered, while the numerical coefficient in front of it  can change.

A conceptually interesting aspect of the result (\ref{rhoH2M2}) is that it appears to involve a mixing of ultraviolet and infrared physics, since it depends both on the UV scale $M$, and  on the horizon size $H^{-1}$,
which represents the ``size of the box'', and therefore plays the role
of an IR cutoff. This is an interesting result by itself, since in quantum field theory we are rather used to the fact that widely separated  energy  scales decouple. Observes that this UV-IR mixing comes out only because our classical subtraction procedure based on \eq{EHH} eliminates the troublesome term 
diverging as $\lc^4$, which is instead a purely UV term. As we already discussed in the Introduction, the origin of this UV-IR mixing can be traced to the fact that, even if $\Tmn(x)$ 
is a local quantity, the vacuum expectation value $\cav\Tmn\vac$ is sensitive to $H(t)$ through
the definition of the vacuum state $\vac$. In fact, the vacuum is  defined from the condition $a_{\bf k}\vac=0$, where
the annihilation operators $a_{\bf k}$ are defined with respect to a set of mode functions $\phi_{\bf k}(t)$;  these mode functions are obtained solving a wave equation over the whole space-time, and therefore are sensitive to the time evolution of the scale factor and, more generally, to the overall geometry of space-time.

\section{Cosmological implications}\label{sect:impl}

\subsection{Vacuum fluctuations and the  dominant component of dark energy\label{sect:domi}}

The results of the previous section show that, with the subtraction that we advocate,
zero-point vacuum fluctuations do not contribute to the cosmological
constant since their energy density, given in \eqs{defOmegaZ}{OmZ}, is not a constant,
but rather a fixed fraction of the critical energy at any epoch.

The first question to be addressed is whether a vacuum energy density with such a time behavior 
could be identified with the dark energy component $\Omega_{\Lambda}$ which is responsible for the observed acceleration of the universe. If this were the case, since we have found that the vacuum  energy density scales as $H^2(t)$,
we would have
$\Omega_{\Lambda}(t)=\Omega_{\Lambda} H^2(t)/H_0^2$ where we 
follow the standard use of denoting by $\Omega_{\Lambda}$ the value of $\Omega_{\Lambda}(t)$ at the present time $t=t_0$.
Such a model has been compared to CMB+BAO+SN data in ref.~\cite{Basilakos:2009wi}, where it is found that it is ruled out at a high significance level. 

Formally the result of
ref.~\cite{Basilakos:2009wi} comes from the fact that the $\chi^2$ obtained fitting this model to the usual estimators for CMB, for BAO and to SNIa turns out to be unacceptably high. However, independently of  technical details, it is easy to understand physically why such a model for dark energy is not viable. Consider in fact a spatially flat model with vacuum energy density $\rho_{\Lambda}(t)$ evolving as $H^2(t)$, with equation of state $w_{\Lambda}=-1$, and matter density 
$\rho_M(t)$, with $\Omega_M+\Omega_{\Lambda}=1$, in the recent universe where we can neglect $\Omega_R$.
The total energy-momentum conservation is
\be\label{totenecon}
\dot{\rho}_M+\dot{\rho}_{\Lambda}+3H\rho_M=0\, ,
\ee
and cannot be split into two separate conservation equations for $\rho_{\Lambda}$ and 
$\rho_M$, $\dot{\rho}_M+3H\rho_M=0$ and $\dot{\rho}_{\Lambda}=0$, since the second equation is obviously incompatible with $\rho_{\Lambda}(t)\sim H^2(t)$, in the recent universe where the Hubble parameter is certainly not a constant. This means that  energy must be transferred between $\rho_{\Lambda}(t)$ and $\rho_M(t)$ in order to obtain the behavior 
$\rho_{\Lambda}(t)\sim H^2(t)$, so we have an interacting dark energy model. The Friedmann equation in this model reads
\bees
\frac{H^2(t)}{H_0^2}&=&\Omega_{\Lambda}(t)+\Omega_M(t)\nn\\
&=&\Omega_{\Lambda}\frac{H^2(t)}{H_0^2}+\Omega_M(t)\, ,
\ees
and therefore
\be\label{Frie1}
\frac{H^2(t)}{H_0^2}=\frac{1}{1-\Omega_{\Lambda}}\, \Omega_M(t)\, 
\ee
and
\be\label{eq43}
\Omega_{\Lambda}(t)=\frac{\Omega_{\Lambda}}{1-\Omega_{\Lambda}}\, \Omega_M(t)\, .
\ee
In this model, therefore, the time evolution of dark energy density is the same as
that of matter. This is clearly incompatible with the existing cosmological observations, that rather indicate that in the recent epoch $\Omega_{\Lambda}(t)$ remained constant, at least to a first approximation, while the matter density in the concordance $\Lambda$CDM model evolves in a way that cannot differ too much from $\Omega_M(t)\sim 1/a^3(t)$. More quantitatively, combining the Friedmann equation
(\ref{Frie1}) with the total energy-momentum conservation equation (\ref{totenecon})
it can be easily found \cite{Basilakos:2009wi} that in this model the dependence of
$\Omega_{M}(t)$ on the scale factor $a(t)$ is
\be\label{OmMa3}
\Omega_{M}(t)=\frac{\Omega_M}{a^{3-3\Omega_{\Lambda}}}\, ,
\ee
where we normalize the scale factor $a(t)$ so that $a(t_0)=1$ at the present epoch $t_0$ (we will see explicitly in 
Sect.~\ref{sect:evolu} how to derive this result in a similar setting).
Physically this result expresses the fact that in this model $\rho_{\Lambda}(t)\sim H^2(t)$ decreases with time, rather than remaining constant as it would do in isolation, and therefore part of its energy density must be transferred to matter, through the energy-conservation equation (\ref{totenecon}). So, matter energy density decreases more slowly that $1/a^3$. For  $\Omega_{\Lambda}(t)$, 
recalling that we are considering a flat model with $\Omega_{\Lambda}+\Omega_M=1$, \eq{eq43} gives
\be\label{Oma3}
\Omega_{\Lambda}(t)=\frac{\Omega_{\Lambda}}{a^{3-3\Omega_{\Lambda}}}\, .
\ee
To compare these results to observations, we recall that
the  possibility of a time evolution of the dark energy density is usually  studied 
in the literature by
parametrizing it as
\be\label{defwDE}
\Omega_{\Lambda}(t)= \frac{\Omega_{\Lambda}}{a^{3+3w_{\Lambda}(t)}}\, . 
\ee
It is important to stress that \eq{defwDE} is simply an effective
way of parametrizing the time dependence of $\rho_{\rm DE}$ in order to compare it with the observations, and the parameter $w_{\Lambda}$ that appears there
is equal to the $w$ parameter that enters in the equation of state of dark energy only if dark energy is not interacting which, as we have seen, is not the case for  a model where $\rho_{\Lambda}(t)\sim H^2(t)$.
 
Comparing \eqs{Oma3}{defwDE} we see  that a model with $\Omega_{\Lambda}(t)=\Omega_{\Lambda} H^2(t)/H_0^2$  predicts that the effective parameter $w_{\Lambda}$ in \eq{defwDE} is constant in time, and equal to $-\Omega_{\Lambda}$.
The limit on constant $w_{\Lambda}$ obtained from  WMAP 7yr data+BAO+SN is
$w_{\Lambda}=-0.980\pm 0.053$
at 68\% c.l. \cite{Komatsu:2010fb} (actually,
this value does not include systematic errors in supernova data; including the systematics, the error on $w_{\Lambda}$ rather becomes about 0.08, see \cite{Amanullah:2010vv}), and therefore reproducing this value would require
$\Omega_{\Lambda}=+0.980\pm 0.053$, and 
$\Omega_M=1-\Omega_{\Lambda}$ consistent with zero to a few percent! It is no wonder that such a model does not fit (any) other  cosmological observation. Alternatively, setting
$\Omega_M\simeq 0.26$ and $\Omega_{\Lambda}\simeq 0.74$, we get
a prediction $w_{\Lambda}\simeq -0.74$, which is excluded  at a 
high confidence level.

Therefore, it appears that the interpretation of zero-point quantum fluctuations as  the dominant dark energy component, responsible for the acceleration of the universe, is not viable. In the next section we will explore the possibility that they could still provide a new, subdominant, ``dark'' component, 
that we will denote by $\rho_{Z}$ to distinguish it from the dominant component $\rho_{\Lambda}$ which instead, according to the limits on $w_{\Lambda}$, is at least approximately constant in time. We will see that, for plausible values of the mass scale $M$ where quantum gravity sets in, the energy density $\rho_Z$  can have a value consistent with existing observations, but 
still potentially detectable.

\subsection{Theoretical expectations for $\Omega_Z$\label{sub:Theo}}

The effect of $\rho_Z$ on the cosmological expansion depends on the mass scale $M$, whose
exact value can only be determined once one has a fundamental theory of quantum
gravity. If we  set $M=\mpl$ and we consider the Standard Model with one
Higgs field, so $n_s=1$,  \eq{OmZ}  gives $|\Omega_Z|\simeq1/(2\pi)\simeq 0.16$. However, precise numerical factors are beyond such
order-of-magnitude estimates  and, by lowering the UV scale $M$,
it is  easy to reduce this number to smaller but still  potentially observable
values. 
For instance, in heterotic string theory the scale is rather given by the
heterotic 
string mass scale $M_H=g\mpl$, 
where $g\simeq 1/5$ is the value of the gauge couplings at the string
scale~\cite{Antoniadis:1999cf}.
This would rather lead to the estimate
$|\Omega_Z|\simeq (n_s+2)g^2/(6\pi)\simeq  2\times 10^{-3}(n_s+2)$.   Lower values of the cutoff, possibly down to the TeV
scale, can be obtained in theories with 
large extra dimensions~\cite{ArkaniHamed:1998nn}. It is also important to observe that,
as explained in the discussion below \eq{rhovac1massiveferm},
the estimate given in \eq{OmZ} holds even in a theory with exact or broken supersymmetry, since the contribution proportional to $H^2M^2$ comes anyhow only from the bosonic sector.

\subsection{Cosmological evolution equations\label{sect:evolu}}

We consider a
flat $\Lambda$CDM cosmology, with a vacuum energy $\Omega_{\Lambda}$ 
having an equation of state 
$p_{\Lambda}=w_{\Lambda}\rho_{\Lambda}$ 
and we further add the energy density $\rho_Z(t)$ given in \eq{defOmegaZ}, with $p_Z=w_Z\rho_Z$. We have nothing to add to the problem of the physical origin of $\rho_{\Lambda}$, except that in our model it is not due to zero-point quantum fluctuations: it has nothing to do with the quartic divergence in the vacuum energy (which is eliminated by our ADM-like subtraction), nor with the quadratic divergence, which is instead the origin of $\rho_Z$.
Then in this model (which could be conveniently called $\Lambda$ZCDM)
 \bees
 \rho&=&\rho_R+\rho_M+\rho_{\Lambda}+\rho_Z\, ,\label{cons2}\\
 p&=&\frac{1}{3}\rho_R+w_{\Lambda}\rho_{\Lambda}+w_Z\rho_Z\, .
 \ees
The values $w_Z=w_{\Lambda}=-1$  will be assumed in the following (in appendix~\ref{App:C} we discuss the case $w_Z$ generic).  
From \eq{defOmegaZ},
\be\label{rhoZH2}
\rho_Z(t)= \Omega_Z\rho_c(t)=
\(\frac{\Omega_Z\rho_0}{H_0^2}\)\, H^2(t)\, ,
\ee
where $\rho_0=3H_0^2/(8\pi G)$ is the present value of the critical density, 
so the energy density in the dark sector is
\bees
\rho_{\rm DE}&\equiv& \rho_{\Lambda}+\rho_Z\nn\\
&=&
\rho_{\Lambda}+\(\frac{3\Omega_Z}{8\pi G}\) \, H^2(t)
\, .\label{rhoZH2b}
\ees
Quite interestingly, this is  the same form of the dark energy density found in 
refs.~\cite{Shapiro:2000dz,Shapiro:2003ui,EspanaBonet:2003vk,Shapiro:2009dh} from an apparently rather different approach,
namely from the suggestion that the cosmological constant could evolve under renormalization group,
after identifying their parameter $\nu$ with our
$\Omega_Z$, compare with eq.~(13) of ref.~\cite{Shapiro:2003ui}. 
Observe also that their value for the parameter $\nu$ is 
$\nu=\pm (1/12\pi) M^2/\mpl^2$ where $M$ is the mass scale where new physics comes in. This has the same parametric dependence on $(M/\mpl)$ as our result for $\Omega_Z$, and is even quite close numerically~\footnote{Observe  that the value
$1/(12\pi)$ of their numerical coefficients depends on the precise definition of the mass scale $M$, which in the approach based on RG involves not only the UV scale, but also some unknown beta-function coefficients, see eqs.~(3.6) and (4.2) of ref.~\cite{EspanaBonet:2003vk}.}.

Using \eq{rhoZH2b}, the Friedmann
equation becomes
\be\label{Friedmod0}
H^2(t)=\frac{8\pi G}{3}
\[\rho_R+\rho_M+\rho_{\Lambda}+\(\frac{3\Omega_Z}{8\pi G}\) H^2(t)\]\, ,
\ee
which can be rewritten as
\be\label{Friedmod}
H^2(t)=\frac{H_0^2}{1-\Omega_Z}
[\Omega_R(t)+\Omega_M(t)+\Omega_{\Lambda}(t)]\, ,
\ee
where $\Omega_i(t)=\rho_i(t)/\rho_0$, ($i=R,M,\Lambda$). Thus, the effect of $\Omega_Z$ on the Friedmann equation is equivalent to a
rescaling of the present value of the Hubble constant, $H_0\ra
H_0/(1-\Omega_Z)^{1/2}$ or, equivalently, to a rescaling of 
 $\Omega_i(t)$ (with $i=R,M,\Lambda$) into $\Omega_i(t)/(1-\Omega_Z)$. Even if we set $w_{\Lambda}=-1$ we have for the moment written $\Omega_{\Lambda}(t)$, rather than setting it to a constant, in order to allow for the possibility of an energy exchange between $\rho_Z(t)$ and $\rho_{\Lambda}(t)$, see below.
 
Zero-point fluctuations also  contribute to the equation for the acceleration 
\be\label{acceq}
\frac{\ddot{a}}{a}=-\frac{4\pi G}{3} (\rho+3p)\, ,
\ee
in a way which depends on the sign of $\rho_Z$. With $w_Z=-1$ (or more generally whenever $w_Z<-1/3$), 
zero-point fluctuations contribute to accelerating the universe if $\rho_Z>0$, while for $\rho_Z<0$ they give a contribution that decelerates the expansion, and which therefore  opposes the accelerating effect of $\rho_{\Lambda}$.

We next consider  the energy conservation equation. As we already mentioned 
in Section~\ref {sect:RDMD},
the fact that zero-point quantum fluctuations
have an energy density $\rho_Z(t)\sim H^2(t)$
has non-trivial consequences on the conservation of energy.
Consider in fact the energy conservation  in a FRW background,
 \be\label{cons1}
 \dot{\rho}=-3H (\rho+p)\, .
 \ee
When there is no exchange of energy among the different components,  energy
conservation is satisfied separately for each component, so $\dot{\rho}_i=-3H
(\rho_i+p_i)$ with $i=R,M,\Lambda,Z$. 
If this were the case, using
 $p_Z=w_Z\rho_Z$, the equation 
 \be
 \dot{\rho}_Z=-3H (\rho_Z+p_Z)=-3(1+w_Z)H\rho_Z
 \ee
 would give $\rho_Z(t)\sim a^{-3(1+w_Z)}$, and in particular
 $\rho_Z(t)$ constant if $w_Z=-1$. However, we have found that 
 $\rho_Z(t)\sim H^2(t)$. So, 
energy must be transferred between zero-point fluctuations
and other components~\footnote{Unless $w_Z$ evolves in time so to track the equation of state of the dominant energy component, i.e. it evolves from $w_Z=1/3$ during RD to $w_Z=0$ during MD. Another interesting possibility, that we do not consider here, is that the Bianchi identities are actually satisfied by assuming a standard conservation law for matter, but assigning a time dependence to Newton's constant, see \cite{Sola:2007sv}.}.

In principle, there are various mechanisms by which vacuum fluctuations can
exchange energy with ordinary matter.  A typical example is the amplification
of vacuum fluctuations \cite{Gri,Star}, or the change in a large-scale scalar
field due to the continuous flow across the horizon of small-scale quantum
fluctuations of the same scalar field, which is also at the basis of stochastic
inflation~\cite{Star2}. If we assume that $\rho_Z$ exchanges energy with $\rho_M$ but not with $\rho_{\Lambda}$ 
the relevant conservation equation is (setting hereafter $w_Z=-1$)
\be\label{rhoMZw0}
\dot{\rho}_M +\dot{\rho}_Z=-3H\rho_M\, .
\ee 
Using \eq {rhoZH2}, we then obtain
\be\label{dotrhow}
\dot{\rho}_M=-3H\rho_M -\(\frac{\Omega_Z\rho_0}{H_0^2}\)\, \frac{dH^2(t)}{dt}\, .
\ee
This equation was already discussed in refs.~\cite{EspanaBonet:2003vk,Basilakos:2009wi} in the context of their time-varying cosmological constant model and, following these papers, to
solve it we  compute the term 
$dH^2(t)/dt$ on the right-hand side by using the Friedmann equation
(\ref{Friedmod}) and we obtain (neglecting for simplicity $\Omega_R(t)$ in the low redshift epoch in which we are  interested here)
\be\label{dotrhow2}
\dot{\rho}_M=-3H\rho_M -\frac{\Omega_Z}{1-\Omega_Z}\dot{\rho}_M\, ,
\ee
where we used the fact that, since we are assuming that $\rho_Z$ only interacts with $\rho_M$, and we are furthermore assuming $w_{\Lambda}=-1$, the energy density $\rho_{\Lambda}$ evolves in isolation and satisfies $\dot{\rho}_{\Lambda}=0$.
\Eq{dotrhow2} can be rewritten as 
\be
\dot{\rho}_M=-3(1-\Omega_Z)H\rho_M\, ,
\ee
and has the  solution 
\be\label{OmegaM(z)eps}
\rho_M(z)=\rho_M(0)(1+z)^{3(1-\Omega_Z)}\, .
\ee
Of course, since we have taken $w_Z=-1$, if vacuum  energy density were non-interacting it would remain constant in time.  In the case $\s=+1$ (i.e. when $\Omega_Z>0$) we have rather found that  $\rho_Z$ is proportional to $+H^2(t)$, and therefore it decreases with time instead of staying constant. This means that dark energy is interacting, and that energy is transferred from vacuum fluctuations to matter, for instance with a mechanism analogous to the amplification of vacuum fluctuations, and as a result the energy density of matter must decrease slower than $1/a^3$. This is  reflected in \eq{OmegaM(z)eps}, since for
$\Omega_Z>0$ we find that $\rho_{M}$ indeed decreases slower than $1/a^3$. If $\Omega_Z<0$ the situation is reversed. A behavior $\rho_Z(t)\sim -H^2(t)$ means that $\rho_Z(t)$ becomes less and less negative as time increases, so energy is transfered from matter to vacuum fluctuations, and $\Omega_M(z)$ decreases faster than $1/a^3$.

In the absence of an understanding of the dynamical origin of the dominant dark energy term $\rho_{\Lambda}$, it is interesting to consider  also the possibility that energy could be exchanged also
between $\rho_Z$ and $\rho_{\Lambda}$.
We assume at first, for simplicity, that energy is exchanged only with $\rho_{\Lambda}$, and not with $\rho_M$. Then the corresponding conservation
equation (taking  $w_{\Lambda}=-1$ for simplicity) is
\be\label{cons}
\dot{\rho}_{\Lambda}+\dot{\rho}_{Z}=0\, ,
\ee
which trivially integrates to 
\be\label{triv}
\rho_{\Lambda}(t)+\rho_Z(t)={\rm constant}=\rho_{\Lambda}(t_0)+\rho_Z(t_0)\, 
\ee
and therefore, using $\rho_Z(t)=\Omega_Z\rho_0 H^2(t)/H_0^2$,
\be\label{OmegaL(z)}
\Omega_{\Lambda}(z)=\Omega_{\Lambda}-\Omega_{Z}\[\frac{H^2(z)}{H_0^2}-1\]\, ,
\ee
where, as usual, on the right-hand side $\Omega_{\Lambda}\equiv \Omega_{\Lambda}(z=0)$.
However, in terms of the total dark energy density 
defined in \eq{rhoZH2b}, we see that in this case we simply have a total dark energy density 
$\rho_{\rm DE}$ that satisfies $\dot{\rho}_{\rm DE}=0$, while the matter energy density satisfies its usual conservation equation $\dot{\rho}_M+3H\rho_M=0$, and also
\eqs{Friedmod0}{acceq}, when rewritten in terms of $\rho_{\rm DE}$, take the standard
$\Lambda$CDM form. Thus,
in the end a model where $\rho_Z$ only exchanges energy with $\rho_{\Lambda}$ (and in which
$w_Z=w_{\Lambda}=-1$)
 is indistinguishable from standard $\Lambda$CDM cosmology with just a cosmological constant. 
In the most general case in which  $w_Z\neq -1$, however, even setting  $w_{\Lambda}=-1$ 
the pressure $p_{\rm DE}=p_Z+p_{\Lambda}=w_Z\rho_Z+w_{\Lambda}p_{\Lambda}$ is no longer equal to $-\rho_{\rm DE}$, so this model has observable deviations for ${\Lambda}$CDM, 
and is in fact of the type called ${\Lambda}
$XCDM~\cite{Grande:2006nn,Grande:2006qi,Grande:2008re,Bauer:2010wj}.
In the following we will however restrict to $w_Z=w_{\Lambda}=-1$.

A more general phenomenological analysis,  in which one takes into account the possibility that $\rho_Z$ interacts both
with $\rho_{\Lambda}$ and with $\rho_M$, can be performed 
by splitting the conservation equation 
\be\label{consgenereal1}
\dot{\rho}_M+\dot{\rho}_{\Lambda}+\dot{\rho}_Z=-3H\rho_M
\ee
into the two equations
\bees
\dot{\rho}_{\Lambda}&=&-(1-\alpha) \dot{\rho}_Z\, ,\label{consgenereal2}\\
\dot{\rho}_M&=&-3H\rho_M-\a\dot{\rho}_Z\label{consgenereal3}\, ,
\ees
where $0\leq\a\leq 1$. \Eq{rhoMZw0} corresponds to the limiting case
$\a=1$ while \eq{cons} corresponds to the limiting case
$\a=0$.  Rewriting these equations in terms of the total dark energy density $\rho_{\rm DE}=\rho_{\Lambda}+\rho_Z$, we get
\bees
\dot{\rho}_{\rm DE}&=&\alpha\dot{\rho}_Z\, ,\label{consgen2}\\
\dot{\rho}_M&=&-3H\rho_M-\a\dot{\rho}_Z\label{consgen3}\, ,
\ees
which shows that the observable consequences depend only on the combination $\a\rho_Z$. In other words, as long as $w_Z=w_{\Lambda}=-1$, there is no point in postulating an  energy exchange between $\rho_Z$ and $\rho_{\Lambda}$, since only the fraction of $\rho_Z$ which is exchanged with $\rho_M$ has observable consequences. In the following we will limit ourselves to
$w_Z=w_{\Lambda}=-1$, and we then set $\a=1$.

\subsection{Limits on $\Omega_Z$ from cosmological observations}\label{sect:comp}

In this section  we perform a first analysis of  the limits that some cosmological
observations impose on $\Omega_Z$. A more detailed comparison with the data will be presented elsewhere. The fact that the energy budget of the universe at the present epoch is known to a precision of about $1\%$ by itself does not yet constraint $\Omega_Z$, since a part of what is normally attributed to $\Omega_{\Lambda}$ could be due to $\Omega_Z$. The only way of disentangling them is by using their different temporal evolution, since $\rho_Z(t)\sim H^2(t)$ while the dominant component
$\rho_{\Lambda}$ is constant, at least within the present experimental accuracy. 
In the next subsections we examine various limits on $\Omega_Z$ which make use of the time dependence $\rho_Z(t)\sim H^2(t)$.

\subsubsection{Bound on $\Omega_Z$ from BBN}

We first examine the  bound coming from  big-bang nucleosynthesis (BBN), which constraints the energy budget of the universe at that  epoch. The limit on extra
contributions
to the energy density at  time of BBN is usually expressed in terms of the
effective number of neutrino species $N_{\nu}$, defined so that 
any extra source of energy density, compared to the Standard Model, is written as
\be\label{boundBBN1}
\frac{\rho_{\rm extra}}{\rho_{\g}}=\frac{7}{8}\Delta N_{\nu}\, , 
\ee
and $\Delta N_{\nu}=N_{\nu}-N_{\nu}^{\rm SM}$, where 
$N_{\nu}^{\rm SM}\simeq 3.046$ is the value predicted by  the Standard Model with three light neutrino families, after taking into account finite temperature QED
corrections and the fact that neutrino decoupling is not instantaneous
\cite{Mangano:2005cc}.  The most recent BBN bound is
$N_{\nu}\leq 3.6$ at 95\% c.l. 
\cite{Iocco:2008va}, corresponding to a limit $\D N_{\nu}\leq 3.6-3.046\simeq
0.55$. 
At the epoch of  BBN only the photons and the three neutrinos contribute significantly to the energy density, while $e^{\pm}$ already annihilated into photons, resulting in a photon temperature higher than the neutrino temperature by a factor 
$(11/4)^{1/3}$. Therefore, at BBN,
\be
\rho_c=\rho_{\g}\[1+3\times \frac{7}{8}\,\(\frac{4}{11}\)^{4/3}\]\, ,
\ee
where the factor $7/8$ comes from Fermi statistics. Combining this with \eq{boundBBN1} gives
\be\label{BoundBBN}
\(\frac{\rho_{\rm extra} }{\rho_c}\)_{\rm BBN}= 
\frac{(7/8)\D N_{\nu}}{1+3\times \frac{7}{8}\,\(\frac{4}{11}\)^{4/3}}\leq
0.29\, ,
\ee
This gives a corresponding bound on the total dark energy density
$\rho_{\rm DE}=\rho_{\Lambda}+\rho_Z$ at the time of nucleosynthesis.
From \eq{consgen2} with $\a=1$ we have
$\dot{\rho}_{\rm DE}=\dot{\rho}_Z$, so
\be\label{evorhoDE}
\rho_{\rm DE}(z)=\rho_Z(z)+ [\rho_{\rm DE}(0)-\rho_Z(0)]\, .
\ee
At $z=z_{\rm BBN}$ the constant term in bracket is totally negligible with respect to the critical density
$\rho_c(z_{\rm BBN})$, while $(\rho_Z/\rho_c)_{\rm BBN}=\Omega_Z$,
so the bound
(\ref{BoundBBN}) translates into
\be
\Omega_Z<0.29\, .
\ee
Strictly speaking, BBN gives an upper bound only on $\Omega_Z$, and not on its absolute value, since negative values of $\rho_Z$ could  in principle be compensated by other  forms of energy.

\subsubsection{Bound on $\Omega_Z$ from CMB+BAO+SNIa}

A comparison of this model to CMB+BAO+SNIa data
has been performed in ref.~\cite{Basilakos:2009wi} within the context of the model
of a cosmological constant that evolves under 
renormalization group~\cite{Shapiro:2000dz,Shapiro:2003ui,EspanaBonet:2003vk,Shapiro:2009dh}. Even if our physical motivations are different from theirs, the model is phenomenologically the same, and our parameter
$\Omega_Z$ corresponds to the parameter that they denote as $\g$ (or as $\nu$). We can therefore translate immediately their result in our setting. In particular, from a joint analysis of CBM+BAO+SNIa, sampling the interval $\Omega_Z\in[0,0.3]$~\footnote{The analysis was actually restricted to positive values of $\Omega_Z$
(J. Sol\`a, personal communication).} in steps of 0.001,
they find a best fit value $\Omega_Z=0.002\pm 0.001$.

\subsubsection{Bound on $\Omega_Z$ from the limits on the time evolution of dark energy}

It is also instructive to  compare  the time evolution 
of a dark energy density of the form
$\rho_{\rm DE}(z)=\rho_{\Lambda}+\rho_Z(z)$ to
the existing observational  limit, which is also obtained from a combination of CMB+BAO+SNIa data.

\begin{figure}
\includegraphics[width=0.45\textwidth]{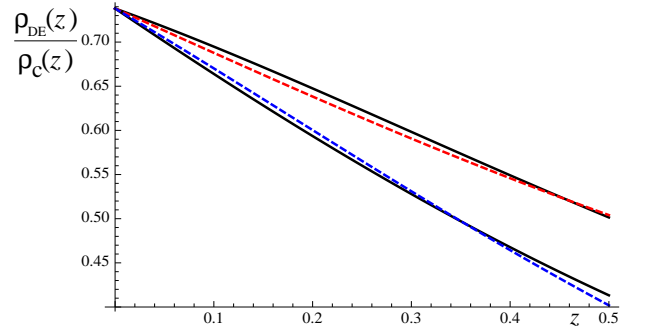}
\caption{\label{fig:boundOZ} The functions
$\rho_{\Lambda}(z)/\rho_c(z)$
for $w_{\Lambda}=(w_{\Lambda})_{\rm min}$ and $(w_{\Lambda})_{\rm max}$  (black solid lines) compared to  the function 
$\rho_{\rm DE}(z)/\rho_c(z)$ given by \eq{rhoDE} with
$\Omega_Z=+0.10$ (upper dashed curve, red) and $\Omega_Z=-0.10$ (lower dashed curve, 
blue).
}
\end{figure}

In standard ${\Lambda}$CDM cosmology,  without the contribution $\rho_Z$, 
the dark energy density is given uniquely by  $\rho_{\Lambda}$ and,
allowing for a generic $w_{\Lambda}$, its evolution with redshift $z$ is given by 
\be
\rho_{\Lambda}(z)=\rho_{\Lambda}(0) (1+z)^{3+3w_{\Lambda}}\, ,
\ee
while (neglecting $\Omega_R$, since  we are interested here in the evolution at small redshifts) the critical density $\rho_c(z)$ is given by
\be
\rho_c(z)=\rho_c(0)[\Omega_M (1+z)^3+\Omega_{\Lambda}]\, ,
\ee
with $\Omega_{\Lambda}\simeq 0.738$. Then
\be\label{ratiorho}
\frac{\rho_{\Lambda}(z)}{\rho_c(z)}=
\frac{\Omega_{\Lambda} (1+z)^{3+3w_{\Lambda}}}{\Omega_M (1+z)^3+\Omega_{\Lambda}}\, .
\ee
We use the recent determination of $w_{\Lambda}$ given in
\cite{Amanullah:2010vv}, 
$w_{\Lambda}=-0.997^{+0.077}_{-0.082}$ at 68\% c.l., which includes also systematic errors on supernova data (the more stringent bound $w_{\Lambda}=-0.980\pm 0.053$ given in
\cite{Komatsu:2010fb} only includes the statistical  error in the SN data). In Fig.~\ref{fig:boundOZ} we then plot
the function $\rho_{\Lambda}(z)/\rho_c(z)$ given in \eq{ratiorho},
in correspondence of the $1\sigma$ upper and lower limits on 
$w_{\Lambda}$, $(w_{\Lambda})_{\rm max}=-0.997+ 0.077$  and 
$(w_{\Lambda})_{\rm min}=-0.997-  0.082$,   respectively
(black solid lines). We limit ourselves to  the redshift interval $0\leq z\leq 0.5$ that, as shown in 
\cite{Kowalski:2008ez,Amanullah:2010vv},  is responsible for most part of the bound on $w_{\Lambda}$. Plotting the constraint on $w_{\Lambda}$, or the corresponding constraints on $\rho_{\Lambda}(z)$, in different redshift bins, one finds in fact that the bin $0.5\leq z\leq 1$ already gives
a poorly constrained $w$, see e.g. Fig.~15 of \cite{Amanullah:2010vv}.

In this range of redshifts, we compare the temporal evolution given in \eq{ratiorho} with
$w_{\Lambda}=(w_{\Lambda})_{\rm min}$  and with 
$w_{\Lambda}=(w_{\Lambda})_{\rm max}$, respectively,
to that obtained in $\Lambda$ZCDM (using for definiteness the values 
$w_{\Lambda}=w_Z=-1$). From \eq{evorhoDE},
\be\label{eq82}
\rho_{\rm DE}(z)=\Omega_Z\rho_c(z)+ \rho_0(\Omega_{\rm DE}-\Omega_Z)\, ,
\ee
where 
$\Omega_{\rm DE}\equiv \Omega_{\Lambda}+\Omega_Z\simeq 0.738$.
To write explicitly the  critical density  $\rho_c(z)=\rho_0H^2(z)/H^2_0$ we
use the Friedmann equation (\ref{Friedmod}) (neglecting again $\Omega_R$) together with
\eq{OmegaM(z)eps}, so
\bees\label{Friedmod2}
\frac{H^2(z)}{H_0^2}&=&\frac{1}{1-\Omega_Z}
[\Omega_M(z)+\Omega_{\Lambda}]\nn\\
&=&\frac{1}{1-\Omega_Z}
[\Omega_M (1+z)^{3(1-\Omega_Z)}+\Omega_{\Lambda}]\, ,
\ees
and therefore, writing $\Omega_{\Lambda}=\Omega_{\rm DE}-\Omega_Z$,

\be\label{eq85}
\rho_c(z)=\rho_0
\frac{\Omega_M (1+z)^{3(1-\Omega_Z)}+\Omega_{\rm DE}-\Omega_Z}{1-\Omega_Z}\, .
\ee
Combining \eqs{eq82}{eq85} we get
\be\label{rhoDE}
\frac{\rho_{\rm DE}(z)}{\rho_c(z)}= \Omega_Z+ 
\frac{(\Omega_{\rm DE}-\Omega_Z) (1-\Omega_Z) }
{\Omega_M (1+z)^{3(1-\Omega_Z)}+\Omega_{\rm DE}-\Omega_Z}\, .
\ee
This function  is plotted in Fig.~\ref{fig:boundOZ},
keeping $\Omega_{\rm DE}$ fixed at the observed  value
$\Omega_{\rm DE}\simeq 0.738$,  and choosing
$\Omega_Z=+0.10$
(upper dashed line, red) and   $\Omega_Z=-0.10$ (lower dashed line, blue). 
We see that, at least at this relatively crude level of analysis,  
values of $\Omega_Z$ in the approximate range
$|\Omega_Z|\,\lsim\, 0.1$
are consistent with 
the observational limits on the temporal evolution of dark energy, since the corresponding curves stay inside  the two curves with $w_{\Lambda}=(w_{\Lambda})_{\rm min}$ and
$w_{\Lambda}=(w_{\Lambda})_{\rm max}$, respectively, down  to  the maximum 
redshifts  $z\simeq 0.5$ where $\rho_{\Lambda}(z)$ is significantly constrained by the data.
If one would rather compare the function $\rho_{\Lambda}(z)/\rho_c(z)$ given in \eq{ratiorho},
to the $3\sigma$ upper and lower limits on 
$(w_{\Lambda})_{\rm max}=-0.997+3\times 0.077$  and 
$(w_{\Lambda})_{\rm min}=-0.997- 3\times 0.082$, one would rather find
$-0.20\lsim \Omega_Z\lsim 0.35$.
Of course this analysis gives only a first rough but intuitive estimate of the bound that can be obtained from the limits on the  redshift dependence of $\rho_{\rm DE}$. A more accurate study requires fitting this model to the data, as in \cite{Basilakos:2009wi}.

\subsection{How not to solve the cosmological constant problem}

We think that another useful aspect of the above analysis is to put
in a sharper focus where  the main difficulty is, in explaining the
observed value of dark energy density. If one  looks at 
\eq{rhoH2M2}, which holds both in RD and in MD, 
 setting $t$ equal to the present time $t_0$ and $M\simeq\mpl$,
one finds that the energy density associated to vacuum fluctuations today is
$\rho_Z(t_0)\sim H_0^2\mpl^2$, which is of the right order of
magnitude of the observed dark energy density (it could even be tempting to
observe,
from \eqs{defOmegaZ}{OmZ}, that with $M=\mpl$, $n_s=12$ and $\s=+1$ one gets
$\Omega_Z\simeq 7/(3\pi)\simeq 0.743$, which is very close to the measured value 
$\Omega_{\Lambda}\simeq 0.738$). However, at this stage this observation is not yet  a possible  explanation of
the numerical value of the cosmological
constant, not even at the level of orders of magnitude. The trouble is that the same computation, performed
at a generic time $t\neq t_0$, gives $\rho_Z(t)\sim H^2(t)\mpl^2$, so the
resulting energy density is not constant.  As we have discussed in Section~\ref{sect:domi}, such a time behavior is observationally excluded, at least for the dominant dark energy component, and can only be accepted for a suitably small subdominant dark  component.

We should observe that the same conclusion  also applies
to some existing
attempts at computing the cosmological constant which make use of
$H_0$ and $\mpl$, such as 
the holographic approach to the cosmological constant
\cite{Cohen:1998zx,Hsu:2004ri,Horvat:2004vn}, 
where  again one obtains
a value  of order $\mpl^2H_0^2$ today. However, the very same reasoning 
would give $\mpl^2H^2(t)$ at a generic time, which as we have seen is ruled out, at least for the dominant component of dark energy.

A similar remark can also be
made for the result of  ref.~\cite{Schutzhold:2002pr}, where it is proposed  that the trace anomaly in QCD gives a
contribution to the vacuum energy density proportional to 
$\Lambda_{\rm QCD}^3$ times the Hubble parameter to the first power. 
Using the present
value of the Hubble parameter,  ref.~\cite{Schutzhold:2002pr} finds that  
$(\Lambda_{\rm QCD}^3H_0)/\rho_0$ is roughly comparable to $\Omega_{\Lambda}$
(actually,  this is true only within about one or two orders of magnitude; for the typical values of
$\Lambda_{\rm QCD}\simeq 100-200\, {\rm MeV}$, we get 
$(\Lambda_{\rm QCD}^3H_0)/\rho_0\simeq 25-200$, not that close to
$\Omega_{\Lambda}=0.7$. Of course precise numerical factor were anyhow beyond the 
estimate in  ref.~\cite{Schutzhold:2002pr}). In any case, the suggestion
of  ref.~\cite{Schutzhold:2002pr}  that this effect has a potential relevance
for explaining  the observed acceleration of the universe faces a problem similar to the one discussed above. In fact, if at the present time $t_0$ one finds
$\rho_{\rm QCD}\sim \Lambda_{\rm QCD}^3H_0$,
the same calculation, performed at a generic time $t$ of
course gives $\rho_{\rm QCD}(t)\sim \Lambda_{\rm QCD}^3H(t)$. As shown in
ref.~\cite{Basilakos:2009wi}, this behavior is ruled out by the comparison with CMB+BAO+SNIa data. (It should also be observed that a contribution to the vacuum energy density proportional to an odd power of $H(t)$ is not consistent with the general covariance of the effective action for gravity, see 
Section~3.1 of ref.~\cite{Shapiro:2008sf}).

What we learn from the above examples is that the real challenge, in explaining the cosmological
constant, is not so much to explain its numerical value today; having
at our disposal the two scales $\mpl$ and $H_0$, once the term
proportional to $\mpl^4$ is eliminated one naturally remains with a
result proportional to $\mpl^2H_0^2$, which gives the right order of
magnitude.
The real challenge is to find a dynamical mechanism that gives
a value of order $\mpl^2H_0^2$ today, without giving $\mpl^2H^2(t)$ at
a generic time $t$, which is the essence of the coincidence problem.

\section{Conclusions}\label{sect:Concl}

One aspect of the cosmological constant problem, or more generally of the problem of understanding the origin of dark energy, is to understand why zero-point fluctuations of quantum fields do not produce an energy density of the order of $M^4$, where $M$ is the UV mass scale  of the quantum field theory (e.g. the Planck mass, or the string mass scale), despite the fact that this seems to be the natural value suggested by quantum field theory. We have proposed that the solutions  to this long-standing puzzle has a purely classical origin, and is related to the correct definition of energy in classical General Relativity, which already involves the subtraction of the flat-space contribution, see \eq{EHH}. 

We have applied this subtraction procedure to a FRW space-time with Hubble parameter $H(t)$ and we have found that the remaining energy density, after renormalization, has a ``natural" value proportional to $M^2H^2(t)$ (and a sign that could in principle be either positive or negative, just as in the Casimir effect). For $M\simeq \mpl$ this gives an energy density just of the order of the critical density of the universe. As we have discussed, however, such an energy density has a time dependence that is not compatible with present observations, if we identify it with the 
dark energy component with $\Omega_{\Lambda}\simeq 0.7$ which  in the standard 
${\Lambda}$CDM cosmology is responsible for the observed acceleration of the universe. It is however possible that it represents a new form of dark energy, whose normalized energy density today, $\Omega_Z$, is smaller than $\Omega_{\Lambda}$. 
Values of $|\Omega_Z|\lsim {\,\,\rm a\,\, few}\times 10^{-3}$  are compatible with the observations that we have discussed, but could  give observable effects in more detailed studies that make use of the specific signature of zero-point fluctuations, namely an energy density with a time dependence proportional to $H^2(t)$, as well as in future more accurate cosmological observations.

\vspace{5mm}
\noindent
{\bf Acknowledgments}. I thank Ruth Durrer,  Juan Garcia-Bellido, Alberto Nicolis,
Massimiliano Rinaldi, Antonio Riotto, Joan Sol\`a and the referees for very
useful comments on the manuscript. This work is supported by the 
Fonds National Suisse.

\appendix

\section{Computation of $\rho$ and $p$ in FRW background \label{app:A}}

Even if the  computation leading to \eqs{rhophik}{pphik} is  elementary, we
find it useful to  report it here. The energy $E$ associated to a real massless scalar
field $\phi$ in a FRW metric, in a comoving volume $V$, is 
\bees
E&=&\int_V d^3x\sqrt{-g} \, T_{00}\nn\\
&=&
\int_V d^3x\sqrt{-g} \, \[\frac{1}{2}(\pa_0\phi)^2+\frac{1}{2a^2}(\pa_i\phi)^2\]
\, ,
\ees
where we used $\Tmn$ from \eq{Tmunu}.
Observe that $x^i$ are comoving coordinates and 
the factor $\sqrt{-g}=a^3$ transforms the comoving volume element $d^3x$ into 
the physical volume element.
Using the mode expansion (\ref{phi})
we get, taking for illustration the term 
$(\pa_0\phi)^2$,
\bees
&&\int d^3x\sqrt{-g} \, (\pa_0\phi)^2=
\int\frac{d^3k}{(2\pi)^3 \sqrt{2k}}\, \int\frac{d^3k'}{(2\pi)^3
\sqrt{2k'}}\nn\\
&&\times\int d^3x\sqrt{-g} \,
\[a_{\bf k}\dot{\phi}_{k} e^{i{\bf k\cdot x}}+
a_{\bf k}^{\dagger}\dot{\phi}^*_{k}e^{-i{\bf k\cdot x}}\]\nn\\
&&\phantom{\int d^3x\sqrt{-g} \,}
\times\[a_{\bf k'}\dot{\phi}_{k'} e^{i{\bf k'\cdot x}}+
a_{\bf k'}^{\dagger}\dot{\phi}^*_{k'}e^{-i{\bf k'\cdot x}}\]\, .\nn
\ees
Performing the integral 
in $d^3x$ over a volume $V$ large compared to the wavelength of all modes of interest we have
\be
\int_V d^3x\,  e^{i({\bf k}\pm{\bf k'})\cdot{\bf x}}= (2\pi)^3\d^{(3)}({\bf k}\pm{\bf k'})\, ,
\ee
and we  get
\bees\label{E1}
E&=&\frac{a^3(t)}{2}\int\frac{d^3k}{(2\pi)^3 {2k}}
\( |\dot{\phi}_k|^2+\frac{k^2}{a^2}|\phi_k|^2\) 
\[ a^{\dagger}_{\bf k}a_{\bf k}+a_{\bf k}a^{\dagger}_{\bf k}\]\nn\\
&&\hspace*{-4.5mm}+\frac{a^3(t)}{2}\int\frac{d^3k}{(2\pi)^3 {2k}}
\[\( \dot{\phi}_k^2+\frac{k^2}{a^2}\phi_k^2\)  a_{\bf k}a_{-\bf k} +{\rm h.c.}\].
\ees
Observe that in flat Minkowski space $a(t)=1$ and (since we are considering a massless field)
$\phi_k(t)\sim e^{-iE_kt}=e^{-ikt}$, so the term proportional to $a_{\bf k}a_{-\bf k}$ vanishes. In a generic curved space it is instead non-zero, so a general state in a curved background is characterized by the expectation values
$\langle a^{\dagger}_{\bf k}a_{\bf k}\rangle$ and 
$\langle a_{\bf k}a_{-\bf k}\rangle$ \cite{Parker:1974qw}. For the vacuum state, however, the only non-vanishing contribution comes from the term proportional to
$a_{\bf k}a^{\dagger}_{\bf k}$ in \eq{E1}, and can be computed using
$[a_{\bf k},a^{\dagger}_{\bf k'}]=(2\pi)^3\d^{(3)}({\bf k}-{\bf k'})$, from which  it
follows that
$[a_{\bf k},a^{\dagger}_{\bf k}]=V$, where $V$ is the comoving spatial volume
(since ${\bf k}$ is a comoving momentum), and therefore
\be
\label{E1bis}
E_{\rm vac}=\frac{1}{2} Va^3(t)\int\frac{d^3k}{(2\pi)^3 {2k}}
\( |\dot{\phi}_k|^2+\frac{k^2}{a^2}|\phi_k|^2\) \, .
\ee
Multiplying the comoving volume $V$ by  the factor
$\sqrt{-g}=a^3(t)$  we recover the physical volume
$V_{\rm phys}$, and therefore the energy of zero-point quantum fluctuations is
\be
E_{\rm vac}=\frac{V_{\rm phys}}{2}\int\frac{d^3k}{(2\pi)^3 {2k}}
\( |\dot{\phi}_k|^2+\frac{k^2}{a^2}|\phi_k|^2\) 
\, .
\ee
The vacuum energy density is then defined as $E_{\rm vac}/V_{\rm phys}$.

For the pressure, the spatial isotropy of the FRW metric implies that
$p=T^1_1=T^2_2=T^3_3$
(observe that, with our signature $(-,+,+,+)$ the energy-momentum tensor of a
perfect fluid is
$T^{\mu}_{\nu}={\rm diag}(-\rho,p,p,p)$). It is convenient to write
$p=(1/3)\sum_i T^i_i$ and, as we have done for the energy density, consider
first the integrated quantity
\bees
P&=&\frac{1}{3}\int_V d^3x\, \sqrt{-g} \,\sum_i T^i_i\\
&=&\int_V d^3x\, \sqrt{-g}\,
\[\frac{1}{2}(\pa_0\phi)^2-\frac{1}{6a^2}(\pa_i\phi)^2\]\, ,\nn
\ees
where, in the second line, the sum over $i$ is understood. Repeating the same
steps as above, we get the zero-point contribution
\be
P_{\rm vac}=
\frac{V_{\rm phys}}{2}\int\frac{d^3k}{(2\pi)^3 {2k}}
\( |\dot{\phi}_k|^2-\frac{k^2}{3a^2}|\phi_k|^2\) 
\, ,
\ee
and $p_{\rm vac}=P_{\rm vac}/V_{\rm phys}$. The off-diagonal elements of  the volume integral of $\Tmn$ vanish trivially,
since they involves integrations over $k_0k_i$, or over $k_ik_j$ with $i\neq
j$, which vanish by parity.

\section{Dependence on the choice of vacuum\label{app:B}}

A point that
deserves some comment is the choice of the modes given in
\eqs{phiketa}{modesRD}. These modes are particularly natural
since in the UV limit they reduce to positive-frequency plane
waves in flat space. However, the choice of the modes is equivalent to the choice
of a particular vacuum state, and the most general possibility is
a superposition of 
positive- and negative-frequency modes (\ref{phiketa}), (\ref{modesRD}) with
Bogoliubov coefficients $\a_{k}$ and $\b_{k}$,
\be\label{phiketaBog}
\phi_k(\eta)=
\frac{\a_{k}}{a(\eta)}\(1-\frac{i\eps}{k\eta}\) e^{-ik\eta}+
\frac{\b_{k}}{a(\eta)}\(1+\frac{i\eps}{k\eta}\) e^{+ik\eta}\, ,
\ee
where $\eps=0$ for RD and $\eps=1$ for
De~Sitter and MD, and the Bogoliubov coefficient satisfy the
normalization condition 
$|\a_{k}|^2-|\b_{k}|^2=1$.
It is straightforward to repeat the computation of the vacuum energy density using the
modes (\ref{phiketaBog}). When computing
$|\phi_k|^2$ and $|\dot{\phi}_k|^2$, mixed term proportional to $\a_k\b^*_k$ 
and to $\a^*_k\b_k$
have a time dependence which contains the factors $\exp\{ \pm 2ik\eta\}$. After integrating over $k$ these produce terms proportional to
$\sin(2a\lc\eta)$ and
$\cos(2a\lc\eta)$. Since
\be
a\lc\eta=\lc a(t)\int^t\frac{dt'}{a(t')}={\cal O}(\lc t)\, ,
\ee
these terms oscillate very fast in time, with a Planckian frequency, and therefore they average to zero over any macroscopic time interval, and can be dropped.
Keeping only the contributions proportional to $|\b_k|^2$ and to 
$|\a_k|^2=1+|\b_k|^2$, subtracting as usual the Minkowski term (and neglecting again the term ${\cal O}(H^4)$ which appears in the MD case), for a real scalar field we find
\be
\rho_{\rm bare}(\lc)=\frac{H^2}{8\pi^2 a^2}\int_0^{a\lc}dk\,	k
(2n_k+1)\, ,
\ee
and $p_{\rm bare}(\lc)=w_{\rm bare}\rho_{\rm bare}(\lc)$, where $w_{\rm bare}$ is the same found before.  Therefore  a different choice
for the vacuum   affects the numerical value of $\Omega_Z$ 
in \eq{defOmegaZ} by a numerical factor which reflects the occupation number of the various modes.

\section{Cosmological equations for $w_Z$ generic\label{App:C}}

In Section~\ref{sect:evolu} we studied the cosmological evolution equations setting $w_Z=-1$. 
In this appendix we  study how vacuum fluctuations affect the matter evolution for $w_Z$ generic.
Then 
\eq{rhoMZw0} generalizes to
\be
\dot{\rho}_M +\dot{\rho}_Z=-3H\rho_M-3(1+w_Z)H\rho_Z\, 
\ee 
and, using \eq {rhoZH2}, we get
\be\label{dotrhowA}
\dot{\rho}_M=-3H\rho_M -\frac{\Omega_Z\rho_0}{H_0^2}\, H [2\dot{H}+3(1+w_Z) H^2]\, .
\ee
The presence of the term proporional to $(1+w_Z)$ on the right-hand side makes it more difficult to find an exact solution. It is however easy to find the solution perturbatively in $\Omega_Z$, which is sufficient for our purposes since we know, from the successes of  $\Lambda$CDM cosmology, that 
$\Omega_Z\ll 1$. Then,
we search for a solution   of the form  
\be
\rho_M(t)=\frac{1}{a^3}[\rho_M(t_0)+\D\rho_M(t)]\, ,
\ee
where, by definition, $\D\rho_M(t_0)=0$ (we set as usual $a(t_0)=1$), and we get
\be\label{dDeltarhoA}
\frac{d}{dt} \D\rho_M=-\frac{\Omega_Z\rho_0}{H_0^2}\, a^3H [2\dot{H}+3(1+w_Z) H^2]\, .
\ee
We solve this equation perturbatively in $\Omega_Z$, so to first order we
simply replace the right-hand side 
of \eq{dDeltarhoA} by its value on the unperturbed  solution
$a(t)=(t/t_0)^{2/3}$ (assuming for simplicity a purely MD phase), with $t_0$ related to $H_0$ by $t_0=2/(3H_0\Omega_M^{1/2})$, and we get
\be
\D\rho_M(t)=-2w_Z\Omega_Z\Omega_M\rho_0\log\frac{t}{t_0}\, ,
\ee
or, in terms of the redshift $z$,
\be
\D\rho_M(z)=3w_Z\Omega_Z\Omega_M\rho_0\log(1+z)\, .
\ee
Therefore 
\be\label{OmegaM(z)A}
\rho_M(z)=\rho_M(0) (1+z)^3\[1+3w_Z\Omega_Z\log(1+z)\]\, .
\ee
This expression is
valid to 
first order in $\Omega_Z$ and, at this order, it is equivalent to
\be\label{OmegaM(z)epsA}
\rho_M(z)=\rho_M(0)(1+z)^{3(1+w_Z\Omega_Z)}\, ,
\ee
which for $w_Z=-1$ agrees with the exact result (\ref{OmegaM(z)eps}). Since the limits on 
$\Omega_Z$ discussed in Section~\ref{sect:impl} basically come from the modified evolution of
$\rho_M$ with red-shift, we see that the limits on $\Omega_Z$ for $w_Z\neq -1$  can be obtained by replacing $\Omega_Z\ra -w_Z\Omega_Z$ in the results of Section~\ref{sect:impl}.


\begin{thebibliography}{99}

\bibitem{Weinberg:1988cp}
  S.~Weinberg,
  Rev.\ Mod.\ Phys.\  {\bf 61} (1989) 1.

\bibitem{Peebles:2002gy}
  P.~J.~E.~Peebles and B.~Ratra,
  Rev.\ Mod.\ Phys.\  {\bf 75}, 559 (2003).

\bibitem{Padmanabhan:2002ji}
  T.~Padmanabhan,
  Phys.\ Rept.\  {\bf 380} (2003) 235.

\bibitem{Copeland:2006wr}
  E.~J.~Copeland, M.~Sami and S.~Tsujikawa,
  Int.\ J.\ Mod.\ Phys.\  D {\bf 15} (2006) 1753.


\bibitem{Maggiore:2005qv}
  M.~Maggiore, ``A Modern Introduction to Quantum Field Theory,''
{\it  Oxford University Press, 2005}, Section~5.7.

\bibitem{ADM} R.~L.~Arnowitt, S.~Deser and C.~W.~Misner (1962).
In {\em Gravitation: an introduction to current research}, L. Witten ed.,
Wiley, New York,
[arXiv:gr-qc/0405109].

\bibitem{pois04} E.~ Poisson, 
``A Relativist's Toolkit. The Mathematics of Black-Hole Mechanics", 
{\it Cambridge University Press,  2004}.

\bibitem{Parker:1974qw}
  L.~Parker and S.~A.~Fulling,
  Phys.\ Rev.\	D {\bf 9} (1974) 341.

\bibitem{Fulling:1974zr}
  S.~A.~Fulling, L.~Parker,
  Annals Phys.\  {\bf 87 } (1974)  176.



\bibitem{Padmanabhan:2004qc}
  T.~Padmanabhan,
  Class.\ Quant.\ Grav.\  {\bf 22} (2005) L107.

\bibitem{Akhmedov:2002ts}
  E.~K.~Akhmedov,
  arXiv:hep-th/0204048.




\bibitem{Barvinsky:1985an}
  A.~O.~Barvinsky and G.~A.~Vilkovisky,
  Phys.\ Rept.\  {\bf 119} (1985) 1.

\bibitem{Buchbinder:1992rb}
  I.~L.~Buchbinder, S.~D.~Odintsov and I.~L.~Shapiro,
  ``Effective action in quantum gravity,''
{\it  Bristol, UK: IOP (1992)}.

\bibitem{Shapiro:2008sf}
  I.~L.~Shapiro,
  Class.\ Quant.\ Grav.\  {\bf 25} (2008) 103001.

\bibitem{Pelinson:2010yr}
  A.~M.~Pelinson and I.~L.~Shapiro,
  arXiv:1005.1313 [hep-th].

\bibitem{Bilic:2010xd}
  N.~Bilic,
  arXiv:1004.4984 [hep-th].



\bibitem{Basilakos:2009wi}
  S.~Basilakos, M.~Plionis and J.~Sol\`a,
  Phys.\ Rev.\  D {\bf 80} (2009) 083511.

\bibitem{Komatsu:2010fb}
  E.~Komatsu {\it et al.}  [WMAP Collaboration],
  Astrophys.\ J.\ Suppl.\  {\bf 192}, 18 (2011).


\bibitem{Amanullah:2010vv}
  R.~Amanullah {\it et al.},
  Astrophys.\ J.\  {\bf 716} (2010) 712.




\bibitem{Antoniadis:1999cf}
  I.~Antoniadis,
  arXiv:hep-th/9909212.

\bibitem{ArkaniHamed:1998nn}
  N.~Arkani-Hamed, S.~Dimopoulos and G.~R.~Dvali,
  Phys.\ Rev.\	D {\bf 59} (1999) 086004.


\bibitem{Shapiro:2000dz}
  I.~L.~Shapiro and J.~Sol\`a,
  JHEP {\bf 0202} (2002) 006.

\bibitem{Shapiro:2003ui}
  I.~L.~Shapiro, J.~Sol\`a, C.~Espana-Bonet and P.~Ruiz-Lapuente,
  Phys.\ Lett.\  B {\bf 574} (2003) 149.

\bibitem{EspanaBonet:2003vk}
  C.~Espana-Bonet, P.~Ruiz-Lapuente, I.~L.~Shapiro and J.~Sol\`a,
  JCAP {\bf 0402} (2004) 006.

\bibitem{Shapiro:2009dh}
  I.~L.~Shapiro and J.~Sol\`a,
  Phys.\ Lett.\  B {\bf 682} (2009) 105.




\bibitem{Sola:2007sv}
  J.~Sol\`a,
  J.\ Phys.\ A {\bf A41 } (2008)  164066.

  
\bibitem{Gri} L. P. Grishchuk, Sov. Phys. JETP {\bf 40} (1975) 409.

\bibitem{Star} A. A. Starobinski, JETP Lett. {\bf 30} (1979) 682.

\bibitem{Star2} A.~A.~Starobinsky, in {\em Field Theory, Quantum Gravity and 
Strings}, eds. H. J. de Vega and N. Sanchez, Springer Verlag
(1986).

\bibitem{Grande:2006nn}
  J.~Grande, J.~Sol\`a and H.~Stefancic,
  JCAP {\bf 0608} (2006) 011.

\bibitem{Grande:2006qi}
  J.~Grande, J.~Sol\`a and H.~Stefancic,
  Phys.\ Lett.\  B {\bf 645}, 236 (2007). 

\bibitem{Grande:2008re}
  J.~Grande, A.~Pelinson and J.~Sol\`a,
  Phys.\ Rev.\  D {\bf 79} (2009) 043006.

\bibitem{Bauer:2010wj}
  F.~Bauer, J.~Sol\`a and H.~Stefancic,
  JCAP {\bf 1012} (2010) 029.


\bibitem{Mangano:2005cc}
  G.~Mangano, G.~Miele, S.~Pastor, T.~Pinto, O.~Pisanti and P.~D.~Serpico,
  Nucl.\ Phys.\  B {\bf 729} (2005) 221.

\bibitem{Iocco:2008va}
  F.~Iocco, G.~Mangano, G.~Miele, O.~Pisanti and P.~D.~Serpico,
  Phys.\ Rept.\  {\bf 472} (2009) 1.

\bibitem{Kowalski:2008ez}
  M.~Kowalski {\it et al.}  [Supernova Cosmology Project Collaboration],
  Astrophys.\ J.\  {\bf 686} (2008) 749.


\bibitem{Cohen:1998zx}
  A.~G.~Cohen, D.~B.~Kaplan and A.~E.~Nelson,
  Phys.\ Rev.\ Lett.\  {\bf 82} (1999) 4971.


\bibitem{Hsu:2004ri}
  S.~D.~H.~Hsu,
  Phys.\ Lett.\  B {\bf 594} (2004) 13.

\bibitem{Horvat:2004vn}
  R.~Horvat,
  Phys.\ Rev.\	D {\bf 70} (2004) 087301.

\bibitem{Schutzhold:2002pr}
  R.~Schutzhold,
  Phys.\ Rev.\ Lett.\  {\bf 89} (2002) 081302.


\end {thebibliography}
\end{document}